\documentclass[fleqn,usenatbib]{mnras}

\usepackage[T1]{fontenc}
\usepackage{ae,aecompl}

\usepackage{graphicx}	
\usepackage{amsmath}	
\usepackage{amssymb}	
\usepackage{xcolor}
\usepackage{cleveref}
\usepackage[normalem]{ulem}

\definecolor{forestgreen}{rgb}{0.13, 0.75, 0.13}

\newcommand\abs[1]{\left|#1\right|}
\newcommand {\etal} {\emph{~et~al.} }

\defcitealias{Noda2019}{N19}
\defcitealias{Nishimichi2018}{N18}
\defcitealias{Beutler2017}{B17}
\defcitealias{Seo2016}{S16}
\defcitealias{Ding2018}{D18}
\newcommand{\noda}{\citetalias{Noda2019}}

\newcommand{\beutler}{\citetalias{Beutler2017}}
\newcommand{\seo}{\citetalias{Seo2016}}
\newcommand{\ding}{\citetalias{Ding2018}}

\newcommand{\hmpc}{\,h\,{\rm Mpc}^{-1}}
\newcommand{\mpch}{\,{\rm Mpc}\, h^{-1}}
\newcommand{\plin}{P_{\rm lin}(k)}

\usepackage{etoolbox}
\makeatletter
\patchcmd\@combinedblfloats{\box\@outputbox}{\unvbox\@outputbox}{}{\errmessage{\noexpand patch failed}}
\makeatother

\title[Barry]{Barry and the BAO Model Comparison}

\author[S. R. Hinton]{
Samuel R. Hinton,$^{1}$\thanks{E-mail: samuelreay@gmail.com}
Cullan Howlett,$^{1}$
Tamara M. Davis$^{1}$
\\
$^{1}$School of Mathematics and Physics, University of Queensland,  Brisbane, QLD 4072, Australia
}

\date{Accepted XXX. Received YYY; in original form ZZZ}

\pubyear{2019}

\begin{document}
\label{firstpage}
\pagerange{\pageref{firstpage}--\pageref{lastpage}}
\maketitle

\begin{abstract}
We compare the performance of four state-of-the-art models for extracting isotropic measurements of the Baryon Acoustic Oscillation (BAO) scale. To do this, we created a new, public, modular code \textsc{Barry}, which contains datasets, model fitting tools, and model implementations incorporating different descriptions of non-linear physics and algorithms for isolating the BAO feature. These are then evaluated for bias, correlation, and fitting strength using mock power spectra and correlation functions developed for the Sloan Digital Sky Survey Data Release 12. Our main findings are as follows: 1) All of the models can recover unbiased constraints when fit to the pre- and post-reconstruction simulations. 2) Models that provide physical descriptions of the damping of the BAO feature (using e.g., standard perturbation or effective-field theory arguments) report smaller errors \textit{on average}, although the distribution of mock $\chi^{2}$ values indicates these are underestimated. 3) Allowing the BAO damping scale to vary can provide tighter constraints for some mocks, but is an artificial improvement that only arises when noise randomly sharpens the BAO peak. 4) Unlike recent claims in the literature when utilising a BAO Extractor technique, we find no improvement in the accuracy of the recovered BAO scale. 5) We implement a procedure for combining all  models into a single consensus result that improves over the standard method without obviously underestimating the uncertainties. Overall, \textsc{Barry} provides a framework for performing the cosmological analyses for upcoming surveys, and for rapidly testing and validating new models. 

\end{abstract}

\begin{keywords}
cosmology -- large-scale structure of Universe -- cosmological parameters
\end{keywords}



\section{Introduction}

Measuring the expansion history of the universe to provide insight into the evolution of the universe and its components is one of the primary focal points of observational cosmology. To do this, various complimentary observational probes have been used to map out the expansion over as wide a redshift range as possible \citep{Weinberg2013}. One of the main probes used is that of Baryon Acoustic Oscillations --- the analysis of the imprint of primordial sound waves from the early Universe \citep{Peebles1970, Sunyaev1970}. 

First detected in the large scale structure (LSS) of galaxy distributions observed with the 2-Degree Field and Sloan Digital Sky Surveys \citep{Cole2005,Eisenstein2005}, BAO analyses now provide some of the best constraints on cosmology with fantastic control of systematic uncertainties 
\citep{Kazin2010,Beutler2011,Blake2011,Anderson2012,Anderson2014,Alam2017}. Modelling of the BAO feature has generally been performed in one of two ways using either the power spectrum $P(k)$ or the correlation function $\xi(s)$. These two methods are related via a Fourier transformation, with consensus measurements often being calculated from modelling measurements of both, then combining these using correlation coefficients evaluated from simulations \citep{Alam2017}. In both of these statistics, we have the choice of extracting information from the isotropic BAO signal \citep{Anderson2012,Beutler2017} by using only the monopole information, or from the anisotropic BAO signal caused by redshift-space distortions (RSD) and the Alcock-Paczynski effect \citep{Alcock1979}. Anisotropic analyses can be performed using higher order multipoles \citep{Anderson2014a,Beutler2017anisotropic} or using angle-dependent wedges \citep{Sanchez2013}, however this latter method has largely been replaced with the standard multipole analysis.

As surveys have increased in statistical power, BAO models have similarly increased in model fidelity and flexibility. Namely, with the inclusion of higher-order perturbation and effective-field theories \citep{Seo2016, Noda2017, Ding2018}, we are now at a point where there are a sufficient number of BAO models that we can investigate the differences between them via direct comparison on simulated data. The above works often perform such a detailed investigation when the new model has been introduced. There have also been some studies using a consistent set of mocks to investigate model assumptions and evaluate systematic bias for particular datasets \citep{VargasMagana2014, VargasMagana2018}. However, in the case of new models, they are usually compared to only one or two existing published results, using idealised simulations (i.e., not full mock galaxy catalogues) or without comparing the new and old models under a consistent modelling framework (i.e., the previous model results are taken directly from the literature and may not be fit in the same way as the new model). For comparisons designed to inform a systematic error budget, a number of other considerations on how the measurements themselves are obtained are often included, leading to difficulty in disentangling the effects of the model physics from other numerical issues or data analysis choices.

In this paper we provide a new comparison of different physical BAO models using a unified fitting treatment and set of simulated data, focusing on which models provide better constraints, the minimum bias, the best computational efficiency, and any preference to fitting to $P(k)$ or $\xi(s)$. We enable this by creating a new, open-source, modular code \textsc{\textsc{Barry}}, which is designed to provide a clear distinction between the various models we test, how the inputs for these models are obtained, how the data is prepared for fitting, and how the posterior samples for each model are obtained. In particular, the code ensures the models remain completely agnostic to different fitting and sampling algorithms, allowing for a robust and clear comparison of their performance. 

We structure the paper as follows: In \Cref{sec:Barry} we introduce \textsc{Barry}. Then in \Cref{sec:models} we provide a general introduction to BAO models, detailing our specific model implementations in \Cref{sec:lit}. We discuss the simulation dataset used in \Cref{sec:dataset},  and compare models in \Cref{sec:pk,sec:xi}. We then discuss procedures to combine the various measurements (for different models and both $P(k)$ and $\xi(s)$) in Section \Cref{sec:consensus}, before concluding in \Cref{sec:conclusion}.

\section{Barry}
\label{sec:Barry}
\textsc{Barry}{\footnote{\url{https://github.com/Samreay/Barry}}} is a python package designed to perform the analysis in this work; and is one of the main outcomes from this study. All results, plots and investigations in this work are produced using this code, and every line of code used within this work (to perform the analysis and make the plots) is publicly available within the code repository\footnote{The only aspect of this entire work we do not provide on the repository, due to the large size of the data, are the full chains for the individual model fits to each mock catalogue used herein (there are a total of 9000 fitting combinations of model and mock). Nonetheless, the repository does contain means and standard deviations of $\alpha$ for each realisation and model, and the full chains are available on request.}. \textsc{Barry}'s philosophy is to remove as many barriers and complications to BAO model fitting as possible and allow each component of the process to remain independent, allowing for detailed comparisons of individual parts as demonstrated herein. To do this, \textsc{Barry} provides:
\begin{itemize}
    \item Standard base model definitions for both the power spectrum and correlation function, to allow for easy extension of these to include more non-linear physics or additional features.
    \item A consistent definition of model parameters, which allows for consistent output of the results of model fitting and the ability to freeze parameters to default values.
    \item A standard dataset definition, with mock datasets produced for the Sloan Digital Sky Survey Main Galaxy Sample \citep{Ross2015} and Data Release 12 \citep{Ross2017, Beutler2017} already included.
    \item Multiple samplers with a single interface to allow hassle-free swapping of posterior sampling methods.
    \item A global fitting class to allow an arbitrary amount of model-dataset pairs to be fit. For instance a single dataset can be fit with multiple models; a single model fit to multiple datasets; or multiple models fit to multiple datasets with ease.
    \item A single entry-point for the application for both local fitting and plotting (i.e., on a laptop or desktop), and for deployment to high-performance computing systems.
    \item Extendable patterns and utilities functions/objects to streamline common tasks such as: \begin{itemize} \item{Pre-generating linear power spectra and higher-order perturbation theory integrals. \textsc{Barry} currently uses \textsc{camb} \citep{Lewis2002,Howlett2012} for this; but other codes such as \textsc{class} \citep{Blas2011} could be easily incorporated.} \item{Converting from power spectra to correlation functions.} \item{Determining a smooth, ``de-wiggled'' power spectrum from a linear power spectrum.} \item{ A standard in-built way of convolving the power spectrum model with the survey window function during model fitting.}\end{itemize} In line with \textsc{Barry}'s philosophy, multiple algorithms for both the Fourier transformation and ``de-wiggling'' are available and can be compared within the code; such a comparison has already been performed in \citet{Hinton2016, Hinton2017}.
\end{itemize}

Additionally, the authors are happy to help any external researcher or group implement their own model variants or future datasets into the software framework if they would find it useful. We aim to try and reduce the amount of time spent rebuilding the wheel, and allow researchers to quickly test model modifications and improvements without significant overhead.

\section{General BAO models}
\label{sec:models}
The vast majority of BAO models begin by taking a generated linear power spectrum, $P_{\rm lin}(k)$, such as one generated by \textsc{camb}, at the effective redshift of the data.

Next, corrections are applied to account for non-linear physics, in particular damping of the BAO feature due to bulk motions \citep{Eisenstein2005}, but also non-linear structure growth, galaxy bias and RSD. Various methods take different approaches. However, common to almost all of them is to separate out the oscillating `wiggle' component of the power spectrum (which encodes the non-linear, biased and redshift-space distorted BAO feature) from the smooth broadband shape (which in addition contains contributions from the non-linear growth of structure, galaxy bias and RSD). Multiple methods exist for this (see \citealt{Hinton2016, Hinton2017} for some comparisons) and usually include obtaining a smooth `de-wiggled' power spectrum denoted $P_{\rm sm}(k)$. Once this is present, the oscillating part can be extracted by simply taking the difference $P_{\rm lin}(k) - P_{\rm sm}(k)$.

By separating out the oscillating and non-oscillating components, we can model different physical processes on each component. Most commonly, given the difficultly in modelling the non-linear power spectrum, the corrections to the smooth power spectrum are modelled with a generic polynomial term to allow for fluctuations in broad shape. Such terms can also simultaneously account for observational systematics. However, this approach is not appropriate for the oscillating power spectrum as we do not wish to artificially wash-out the BAO feature we are trying to measure. Non-linear modifications to this term are often written in terms of the `propagator', which determines the evolution of the density field from early time to late time. Some methods provide simple parameterisations of the propagator, where the non-linear motion of galaxies dampening the BAO feature at small scales (higher $k$) are included using only an exponential damping function with a free parameter. Other models include higher-order contributions from Standard Perturbation Theory (SPT), Effective Field Theory (EFT) or Langrangian Perturbation Theory (LPT) to try and provide a better and more physically motivated propagator. In these cases the damping of the `wiggle' power spectrum still takes the form of an exponential decay (which motivates the form of the simpler model), but with less freedom in the decay rate.

In order to make model comparison easier, we standardise the definition of various models, such that a generic model takes the form 
\begin{align}
    P(k) = \int_{-1}^1 d\mu \left[\mathcal{A} + \mathcal{B} P_{\rm sm}(k) \right]\left[1 + \left(\frac{P_{\rm lin}(k)}{P_{\rm sm}(k)} - 1\right) \mathcal{C}^2 \right] + \mathcal{D}, \label{eq:model}
\end{align}
where $P_{\rm lin}$ denotes the linear power spectrum, $P_{\rm sm}$ denotes a de-wiggled, smooth power spectrum, and $\mu$ is the cosine to the line-of-sight angle. The prefactor $\mathcal{B}$ is some function applied to the smooth power spectrum, such as a linear bias correction, Kaiser factor \citep{Kaiser1987} or Fingers-of-God damping term, $\mathcal{C}^2$ is our propagator that determines the evolution of the BAO feature, and $\mathcal{A}$ and $\mathcal{D}$ are some extra additive terms, which are set to polynomial functions for many models. We choose to write out our models in this way to explicitly separate out the oscillatory BAO signal from the broadband shape. 

Whilst we include the integral over $\mu$ in the general form of the model, we investigate only isotropic models in this work, where we are integrating over $\mu$ and taking the monopole term. Given our model definition, inclusion of anisotropies by calculating the quadrupole and hexadecapole through adding in the relevant Legendre polynomials is hence straightforward, and will be implemented in future work.
 
\subsection{Reconstruction}

In this work we investigate models with and without reconstruction \citep{Eisenstein2007,Kazin2010}, where reconstruction is the technique of using the density field to infer rough galaxy motions, allowing galaxies to be moved back to a position closer to their initial, early time location. This linearises the BAO signal in the data and (on average) strengthens it compared to the smooth clustering, making it easier to isolate and fit. As the observed density field contains shot-noise and is only measured at late times, reconstruction is imperfect. However it still offers significant gain in BAO signal, typically improving BAO constraints by a factor of $20\%$-$50\%$ \citep{Padmanabhan2012,Anderson2012,Burden2014,Kazin2014}.

Where appropriate, we have assumed that reconstruction has been performed using the `isotropic convention' \citep{Padmanabhan2012, Anderson2014}, which attempts to remove the RSD signal during the reconstruction process by dividing the observed density field by the Kaiser factor during the computation of the displacement and without adding this factor back on to the displaced galaxies or random points. This leads to a slightly more linear BAO feature compared to other methods, at the cost of an overall reduction in clustering amplitude \citep{Schmittfull2015,White2015,Seo2016,Chen2019}. We also assume that a Gaussian smoothing of the density field was used and define the smoothing kernel in Fourier space as
\begin{align}
    S(k) = \exp\left[-\frac{1}{2} k^2 \Sigma_{\rm smooth}^2\right],
\label{eq:smoothing}
\end{align}
mirroring that of \citet{Anderson2012, Beutler2017, Noda2019}. This kernel is used to smooth out the density field during reconstruction, and the length scale $\Sigma_{\rm smooth}$ needs to strike a balance in size --- too small and shot noise results in a noisy estimate of the density field, too large and the estimated displacement is overly suppressed, resulting in a poor reconstruction. For convenience, we denote $S_c(k) \equiv 1 - S(k)$ the complimentary smoothing kernel. We note that some papers define the kernel using a factor of $\frac{1}{4}$ \citep{Seo2016, Ding2018} instead of $\frac{1}{2}$; for consistency all the models in this work have been standardised to a fraction of a half. 

\subsection{BAO dilation parameter}

In all cases, we fit for the BAO feature using the standard $\alpha$ parameter \citep{Eisenstein2007,Anderson2012}. Analysis of galaxy clustering requires one to assume a fiducial distance cosmology to convert galaxy redshifts and sky positions to Cartesian coordinates. Differences between the fiducial distance cosmology used in this process and the fiducial simulation cosmology of the data (or the cosmology used to generate a simulation) manifest as a dilation of scales in the power spectrum, changing the wavelength of the BAO oscillations (and the peak position in configuration space). The BAO scale can then be extracted by fitting a template for the power spectrum or correlation function with a free parameter $\alpha$ that models this dilation and shifts the template as a function of scale. The best-fit value of $\alpha$ encodes the difference between the BAO scale in the fiducial model (fid) and true cosmologies via
\begin{equation}
    \alpha = \biggl(\frac{D_{A}(z)}{D_{A}^{\mathrm{fid}}(z)}\biggl)^{2/3}\biggl(\frac{H^{\mathrm{fid}}(z)}{H(z)}\biggl)^{1/3}\biggl(\frac{r_{s}^{\mathrm{fid}}}{r_{s}}\biggl)
\end{equation}
where $D_{A}(z)$ and $H(z)$ are the angular diameter distance and Hubble parameter at the effective redshift of the galaxy sample respectively, and $r_{s}$ is the radius of the sound horizon at the baryon drag epoch. Hence, measurements of the BAO dilation parameter provide cosmological information through the dependence of these three parameters on the underlying cosmological model. Anisotropic BAO measurements allow one to go even further and break the degeneracy between $D_{A}(z)$ and $H(z)$.

The fitting for the power spectrum is performed by evaluating the model for the $i^{th}$ measurement bin at $k = k_i/\alpha$ and then comparing to the measurement at $k_i$. For the correlation function it is performed by evaluating the model for the $i^{th}$ measurement bin at $s = s_i \alpha$ and then comparing to the measurement at $s_i$; 
\begin{align}
    P^{\mathrm{data}}(k_{i}) &= P^{\mathrm{model}}(k_{i}/\alpha) \notag \\
    \xi^{\mathrm{data}}(s_{i}) &= \xi^{\mathrm{model}}(s_{i}\alpha).
\end{align}
The dilation parameter $\alpha$ also impacts the volume element and thus the overall power in each bin, however this is absorbed by parameters marginalising the amplitude of the signal, typically the linear bias $b$, and does not need further correction.

\subsection{Fiducial Cosmologies}

It is important to keep our nomenclature clear on what we mean by fiducial cosmology, for we have three different cosmologies going into this analysis.

\begin{enumerate}
    \item Fiducial simulation cosmology: The cosmology used to generate the mock simulations.
    \item Fiducial dataset cosmology: The cosmology used to convert the mock simulation redshifts to distances.
    \item Fiducial model cosmology: The base cosmology used in model fitting to generate a linear power spectrum.
\end{enumerate}

As highlighted below in \Cref{sec:dataset}, the fiducial cosmologies for (i) and (ii) differ in our chosen dataset, as they likely would in the real Universe where the cosmology used to convert the galaxy redshifts is not the same as the Universe's true cosmology. However, it is common, as is done herein, to assume the same cosmology for (ii) and (iii). \cite{Xu2012, VargasMagana2014} have investigated how differences between (i) and (ii)=(iii), and (i)=(ii) and (iii) change or bias the resulting value of $\alpha$. They found that the change in $\alpha$ when (i) and (ii) differ agrees with the expectation to within 0.001, and that the change in $\alpha$ using different fiducial model cosmologies is also less than 0.001. An interesting piece of future work would be to perturb all three choices of fiducial cosmology within \textsc{Barry} to determine whether models with more cosmology dependence lead to higher bias when the cosmologies differ. However, given the small change in $\alpha$ from the above references we would expect such a comparison to be more useful or conclusive with anisotropic fits. As such we leave this test for future work.

\section{Methods in Literature}
\label{sec:lit}

Here we give an overview of the models implemented inside \textsc{Barry} and tested in this paper. Models are referenced using the study they are based on. We note that the models may not be an exact replication of the original works and may have been modified slightly to match our general model parameterisation given in Eq.~\ref{eq:model}. In these cases such differences are highlighted below, and the key features and motivations of each model are still retained. This list should be treated as representative, but not exhaustive, and there are a number of ways each of the models below could be modified slightly (some of which has been tested in the course of this work). The main motivation of \textsc{Barry} is to allow one to test such modifications, ideally beyond that shown even in this work.

\subsection{Beutler\etal (2017)}
Methods used in the SDSS collaboration provide a natural model to implement, as the evolution of the model from SDSS-II to SDSS-IV means that many model choices have been systematically evaluated over repeated implementations and testing. The model presented in \citet[][hereafter denoted {\beutler}]{Beutler2017} is similar to the model used in \citet{Anderson2014}, with the addition of a Fingers-of-God term to account for non-linear redshift space distortions (which should also somewhat be accounted for by the polynomial terms used for the broadband power spectrum shape). We implement this model using
\begin{align}
    \mathcal{B} &= B^2 \left(1 + \frac{1}{2} k^2 \Sigma_s^2\right)^{-2}, \\
    \mathcal{C}^2 &= \exp\left[-\frac{1}{2}k^2\Sigma_{nl}^2\right],
\end{align}
where $B$ is the bias, $\Sigma_s^2$ is the damping scale for the Fingers-of-God, and $\Sigma_{nl}$ allows for a smooth dampening of the BAO feature on smaller scales. In the case of reconstruction, $B$ represents primarily the linear bias, with it also absorbing terms from RSD in the pre-reconstruction case. This freedom in the propagator means that the {\beutler} method does not depend on $S(k)$. We also set
\begin{align}
    \mathcal{A} = \begin{cases}
    a_1 k + a_2 + \frac{a_3}{k} + {a_4}{k^2} + {a_5}{k^3},&\text{ pre-recon} \\
    a_1 k^2 + a_2 + \frac{a_3}{k} + {a_4}{k^2} + {a_5}{k^3},&\text{ post-recon,}
    \end{cases}
\end{align}
and $\mathcal{D} = 0$. The five polynomial terms allow enough flexibility to model the broadband shape of the power spectrum, with {\beutler} reporting a preference for the $k^2$ term over $k$ for their post-reconstruction fits. This gives the isotropic model defined in {\beutler} a total of 9 free parameters: $\alpha$, $B$, $\Sigma_{nl}$, $\Sigma_s$, $a_{1\dots5}$. In the various studies using this or similar a model (i.e., \citealt{Anderson2014,Ross2015,Beutler2017}), it has been common to apply a reasonably tight prior on $\Sigma_{nl}$ based on fits to mock surveys; it is often fixed or has a Gaussian prior with width $\sim 2h^{-1}\mathrm{Mpc}$. In this work we test the {\beutler} model with free and fixed $\Sigma_{nl}$ values for completeness. We note that our model implementation actually differs slightly from that described in {\beutler}, which would assign the polynomial terms to $\mathcal{D}$ instead of $\mathcal{A}$. Setting the polynomial terms to $\mathcal{D}$ resulted in a weak BAO feature even when fixing $\Sigma_{nl}=0$ and so we move them into $\mathcal{A}$ (which was also the way the terms were included in \citealt{Anderson2014,Ross2015}). Finally, we note that when implementing the correlation function model, we remove the Fingers-of-God term, to bring the model into line with the implementations of \citet{Anderson2014,Ross2015,Ross2017}.

\subsection{Seo\etal (2016)}

The second model we implement in our comparisons is based on the work of \citet[][hereafter denoted {\seo}]{Seo2016}. {\seo} follow \citet{Matsubara2008}, constructing a model of the pre- and post-reconstruction power spectrum using Lagrangian perturbation theory (LPT) up to second-order. In their original work, {\seo} then use this to motivate a simple anisotropic model for the damping of the BAO feature pre- and post-reconstruction, which takes the form of an exponential similar to that used in \cite{Beutler2017}. In fact the models of \cite{Beutler2017} as given above are a result of the {\seo} work;  the anisotropy is simply removed given that we are only interested in fitting the monopole of the clustering. To test the impact of including a more physically motivated treatment of the BAO damping, here we reproduce the LPT model for the oscillatory part of the power spectrum in full --- including terms deemed negligible by {\seo}, and without allowing the BAO damping terms to vary or differ from their LPT predictions. However, in order to account for the poor performance of perturbation theory on non-linear scales, we do not adopt LPT for the smooth power spectrum component.

As such, we keep $\mathcal{A}$, $\mathcal{B}$ and $\mathcal{D}$ identical to the {\beutler} model, and modify $\mathcal{C}$ such that for post-reconstruction models we have
\begin{align}
    \mathcal{C} &= \left[1 + \beta \mu^2 S_c(k) \right] \exp\left[-\frac{1}{2}k^2\Sigma_{\rm dd}^2 \phi \right] \notag \\
    &\quad + \frac{S(k)}{B}\left[e^{-k^2\Sigma_{\rm ss}^2/2} - e^{-\frac{1}{2}k^2\Sigma_{\rm dd}^2 \phi} \right] \label{eq:seoprop}\\
    \phi &= \left\lbrace 1 + (2+f)f\mu^2 \right\rbrace \\
    \Sigma_{\rm dd}^2 &= \int \frac{dq}{6\pi^2} P_{\rm lin}(q) S_c^2(q) \\
    \Sigma_{\rm ss}^2 &= \int \frac{dq}{6\pi^2} P_{\rm lin}(q) S^2(q),
\end{align}
where $\beta = f/B$ and $f$ is the growth rate of structure, which we fix to $\Omega_m^{0.55}$ (although this can be treated as a completely independent parameter in \textsc{Barry}). This model is only up to first-order in LPT consistent with the results of {\seo} and the expectation that reconstruction linearises the BAO feature and removes non-linearities that would be modelled with higher-order theory. For pre-reconstruction models, we refer to Appendix B of {\seo} which uses second-order LPT. Setting $S(k) = 0$ (as appropriate for pre-reconstruction), which removes or simplifies a number of the perturbation theory terms, we have
\begin{align}
    \mathcal{C} &= \left\lbrace1 + \beta\mu^2 + \frac{3}{7}\left(\frac{R_1(k)}{\plin} + \frac{R_2(k)}{\plin} \right)\left(1 + 2f\mu^2\right) \right.\notag\\
    &\left.-\frac{3}{7}\frac{R_1(k)}{\plin}\left(\frac{4}{9B} + \frac{1}{3}\beta\mu^2 \right) \right\rbrace e^{-\frac{1}{2}k^2\Sigma_{\rm dd}^2 \phi}
\end{align}
with $R_1(k)$ and $R_2(k)$ defined from \citet{Matsubara2008} as
\begin{align}
    R_1(k) &= \frac{k^3}{4\pi^2} \plin \int_0^\infty dr P_{\rm lin}(kr)\times \notag \\
    &\quad\left[-\frac{1+r^2}{24r^2}\left(3-14r^2+3r^4\right) + \frac{(r^2-1)^4}{16r^3}\ln\abs{\frac{1+r}{1-r}}\right] \\
    R_2(k) &= \frac{k^3}{4\pi^2} \plin \int_0^\infty dr P_{\rm lin}(kr)\times \notag \\
    &\quad\left[-\frac{1+r^2}{24r^2}\left(3-2r^2+4r^4\right) + \frac{(r^2-1)^3(1+r^2)}{16r^3}\ln\abs{\frac{1+r}{1-r}}\right].
\end{align}
In the {\seo} definition of their model (their Equation 25), the propagator is simplified and the propagator terms are applied to both the wiggle component and the smooth power spectrum. However, as we are only modelling the monopole, the $\mu$ terms integrate out and we are left with only a bias factor affecting the smooth power spectrum, keeping our model definition in Eq.~\eqref{eq:model} consistent. This formulation gives us 8 free parameters: $\alpha$, $B$, $a_{1\dots5}$ and $\Sigma_s$.

\subsection{Ding\etal (2018)}

Our third model is inspired by the EFT0 model from \citet[][hereafter denoted {\ding}]{Ding2018}. The EFT0 and EFT1 models (based on effective field theory arguments) presented in {\ding} include scale-dependent bias in the BAO feature, but neglect other higher order terms and only consider the oscillatory part of the power spectrum; the model does not give a prediction for the smooth component of the power spectrum. We implemented a model derived from the EFT0 for simplicity, as the EFT1 model is different for pre- and post-reconstruction. However, incorporating and testing an EFT1-style model would be easy given \textsc{Barry}'s architecture. As {\ding} only model the oscillatory part of the power spectrum, for consistency, we adopt the same smooth component as used in {\beutler} and {\seo}. Hence, we have the same linear bias and Fingers-of-God terms from previous models (keeping $\mathcal{A}$, $\mathcal{B}$, $\mathcal{D}$ as defined above), but update the propagator:
\begin{align}
    \mathcal{C}^2 &= \left\lbrace \left[1 - \frac{S(k)}{B} + \beta\mu^2S_c(k) + \frac{b_\delta k^2}{2Bk_L^2} \right]^2 - \left[\frac{b_\delta k^2}{2Bk_L^2} \right]^2 \right\rbrace\times \notag\\
    &\quad e^{-k^2(1+(2+f)f\mu^2)\Sigma^2_{dd,nl}} + \frac{S^2(k)}{B^2}e^{-k^2\Sigma^2_{ss,nl}} \notag\\
    &\quad - 2\left[1 - \frac{S(k)}{B} + \beta\mu^2S_c(k) + \frac{b_\delta k^2}{2Bk_L^2}\right] \frac{S(k)}{B}e^{-k^2(1+f\mu^2)\Sigma^2_{sd,nl}}
\end{align}
where $k_L = 1 \hmpc$ and
\begin{align}
    \Sigma^2_{dd,nl} &= \int \frac{dq}{6\pi^2} P_{\rm lin}(q) S_c(q)(1 - j_0(r_s q)), \\
    \Sigma^2_{sd,nl} &= \int \frac{dq}{6\pi^2} P_{\rm lin}(q)\left[\frac{S^2(q)+ S_c^2(q)}{2}  - j_0(r_s q)S_c(q)S(q) \right], \\
    \Sigma^2_{ss,nl} &= \int \frac{dq}{6\pi^2} P_{\rm lin}(q) S^2(q) (1 - j_0(r_s q)).
\end{align}
$j_0$ is the $l=0$ spherical Bessel function of the first kind, and $r_s$ is the radius of the sound horizon at the baryon-drag epoch for the model cosmology, giving the BAO scale. Fits with varying cosmology will therefore have varying $r_s$ calculated from the model cosmology, we do not fix it to a specific value. For pre-reconstruction $S(k) = 0$, reducing the propagator down to 
\begin{align}
    \mathcal{C}^2 &= \left\lbrace \left[ 1 + \beta\mu^2 + \frac{b_\delta k^2}{2Bk_L^2} \right]^2 - \left[\frac{b_\delta k^2}{2Bk_L^2}\right]^2 \right\rbrace e^{-k^2(1+(2+f)f\mu^2)\Sigma_{dd,nl}^2}.
\end{align}
 Following {\ding}, we fix $k_L = 1\hmpc$. Unlike most model parameters in \textsc{Barry}, we do not actually give $k_{L}$ the potential to vary as it is completely degenerate with $b_{\delta}$ in the EFT-based models. This formulation gives us 9 free parameters: $\alpha$, $B$, $b_\delta$, $a_{1\dots5}$ and $\Sigma_s$.
 
\subsection{Noda\etal (2019)}

The final model we investigate in this paper is that of \citet[][hereafter denoted {\noda}]{Noda2019}. Their work combines the derivation of a BAO extraction algorithm to separate out the acoustic oscillations from the broadband power spectrum, and the application of a BAO model onto a mixture of the extracted and non-extracted power spectrum. To begin, \citet{Noda2017, Nishimichi2018, Noda2019} lay out the definitions of an algorithm to extract the BAO signal. To summarise, for binned (discrete) data, the extractor is given by
\begin{align}
    R(k_i, \Delta) = \frac{\sum_j f_{ij} \left(1 - \frac{P(k_j)}{P(k_i)} \right) }{ \sum_j f_{ij} \left[ 1 - \cos \left(r_s(k_j - k_i) \right) \right]}, \label{eq:extractor}
\end{align}
where
\begin{align}
    f_{ij} &= \begin{cases}1&\quad \abs{k_j - k_i} \leq 2\pi\Delta/r_s, \\ 0&\quad\text{otherwise},  \end{cases}
\end{align}
and where we follow {\noda} and set $\Delta=0.6$. Numerically, $f_{ij}$ is simply a window function defining over how many bins the filter applies.  This filter function, converting a binned power spectrum $P(k)$ to $R(k,\Delta)$ is what we term the `BAO Extractor' in the rest of the paper, and can be applied onto any model, so long as it is also applied onto the dataset with a consistent choice of $\Delta$. The model used by {\noda} differs considerably from previous models in that it attempts to model the non-linear oscillatory \textit{and} smooth parts of the power spectrum. Rewriting Equation 9 from {\noda} to more closely resemble our other models, we have
\begin{align}
    \mathcal{A} &= 0, \\
    \mathcal{B} &= B^2 e^{-Ak^2} \left[1 + \beta\mu^2 S_c(k) \right]^2, \label{eq:noda1}\\
    \mathcal{C} &=  e^{-k^2 \Sigma^2_{rs}(\mu) / \gamma_{rec}}, \\
    \mathcal{D} &= B^2 e^{-Ak^2} \Delta P_{sm,nl}(k,\mu)\label{eq:noda2},
\end{align}
where
\begin{align}
    \Sigma_{rs} &= \frac{1 + f\mu^2(2+f)}{6\pi^2} \int_0^\infty dq P_{\rm sm}(q) \left[ 1-j_0(qr_s) + 2j_2(qr_s)\right] \notag\\
    &\quad + \frac{f\mu^2(\mu^2-1)}{2\pi^2}\int_0^\infty dq P_{\rm sm}(q) j_2(qr_s)
\end{align}
describes the damping of the BAO feature, and
\begin{align}
    \Delta P_{sm,nl}(k,\mu) = \Delta P_{sm,\delta\delta}(k)  + 2\beta\mu^2\Delta P_{sm,\delta\theta}(k) + \beta^2\mu^4 \Delta P_{sm,\theta\theta}
\end{align}
describes the non-linear corrections to the smooth component of the redshift-space power spectrum. $\Delta P_{sm,\delta\delta}$, $\Delta P_{sm,\delta \theta}$, and $\Delta P_{sm,\theta\theta}$ arise from the matter ($\delta$) and velocity divergence ($\theta$) auto- and cross-power spectra computed using the smooth linear power spectrum. These can be obtained a number of ways; {\noda} use standard perturbation theory, the expressions for which have been reproduced a number of times in the literature and so will not be repeated here (see e.g., \citealt{Makino1992,Scoccimarro2004,Vlah2013}, or \citealt{Nishimichi2007} for a particularly concise presentation).

In this work, we also consider an alternative calculation of the non-linear correction $\Delta P_{sm,nl}(k,\mu)$ using the inplementation of the `halofit' matter power spectrum \citep{Smith2003,Takahashi2012,Mead2016} included in \textsc{camb} and the fitting formulae for the power spectrum involving the velocity divergence from \cite{Jennings2012a,Jennings2012b}. A comparison of the two models in given in \Cref{sec:noda_nonlinear}.
 
 As for other models, by default we fix $f=\Omega_m^{0.55}$, giving the {\noda} model only three free parameters: $\alpha$, $B$ and $A$. Following {\noda} we fix $\gamma_{rec} = 4$ for post-reconstruction and $\gamma_{rec} = 1$ pre-reconstruction. {\noda} also fixes $B$ based on an initial fit to just the power spectrum up to $k\leq0.15 \hmpc$ but we do not do this by default. However, it is worth noting that \textsc{Barry} allows for parameters to be fixed or freed extremely easily and we will explore the consequences of fixing both $\gamma_{rec}$ and $B$ in this work. Overall, given its physically motivated smooth-model and fixed determination of $\Sigma_{rs}$, the {\noda} model contains the fewest free parameters of all the models we test.

\subsection{Correlation function models}

In addition to testing the power spectrum version of the models above, we also test their efficiency in fitting to the correlation function. In general, this involves taking the model $P(k)$ \textit{without adding on any polynomial terms}, applying the Spherical Hankel transformation, and then including the polynomial terms afterwards. Mathematically, for the {\beutler}, {\seo} and {\ding} models
\begin{equation}
    \xi(s) = \frac{1}{2\pi^{2}}\int_{0}^{\infty}k^{2}P(k; \mathcal{A}=0,\mathcal{D}=0)j_{0}(ks)dk + \mathcal{E} 
\end{equation}
where $j_{0}(ks)$ is the zeroth-order spherical bessel function, and we base our polynomial terms on \cite{Ross2015, Ross2017}
\begin{equation}
    \mathcal{E}=\frac{a_{1}}{s^{2}} + \frac{a_{2}}{s} + a_{3},
\end{equation}
who found from a $\chi^2$ robustness test that fewer polynomial terms were needed for the correlation function fits than power spectrum fits. The correlation function models hence contain 2 fewer free parameters in general than their Fourier-space counterparts.

The {\noda} model is designed mainly to work with the BAO extractor laid out in \cref{eq:extractor}, which does not have an easy analogy for the correlation function. As such we do not test the {\noda} model on the correlation function (although the power spectrum model underlying {\noda} could be incorporated). As such, we have a total of nine models, five for the power spectrum (with {\beutler} counting twice for when we fix or do not fix $\Sigma_{nl}$) and four for the correlation function.

\section{Comparison Datasets}
\label{sec:dataset}

In this paper, we present results based on the MultiDarkPATCHY mock-simulations \citep{Kitaura2016} created for the Baryon Oscillation Spectroscopic Survey \citep[][BOSS]{Dawson2013}, which forms part of the Sloan Digital Sky Survey III \citep[][SDSS-III]{Eisenstein2011}. To simplify our analysis, we utilise only the North Galactic Cap (NGC) data for the highest redshift bin $0.5<z<0.75$ (effective redshift $z=0.61$) which represents the highest statistics (number of objects) for a single dataset we could find with both $P(k)$ and $\xi(s)$ measurements available for public use. We analyse the two sets of measurements independently and only combine $\alpha$ values later, following their treatment in the literature. A joint fit of both sets of measurements could be difficult due to the potentially different systematics in each and their high degree of correlation leading to a covariance matrix that is difficult to invert. Nonetheless, we note that this is a possible way of extending the analysis and \textsc{Barry} in the future. The MultiDarkPATCHY mocks detailed in \citet{Kitaura2016} provide us with a thousand mock simulations for both pre- and post-reconstruction of the BAO signal, for both $P(k)$ and $\xi(s)$. The fiducial cosmology in our analysis of these mocks is the Flat $\Lambda$CDM cosmology parameterised by $\Omega_m=0.31$, $h_0=0.676$, $\Omega_b = 0.04814$, $n_s=0.97$, which matches that used in \cite{Beutler2017,Ross2017} to make the clustering measurements\footnote{We actually use a different value for $n_{s}$ compared to \cite{Beutler2017,Ross2017} (wherein $n_{s}=0.96$), however this has no effect on the measurements because the conversion from angular to cartesian coordinates does not depend on $n_{s}$.}. We note that this does \textit{not} match the underlying cosmology of the simulations used to create the mocks and as such we would expect to recover $\alpha=0.9982$. This is the value used in all subsequent sections as our expectation value of $\alpha$.  The reconstruction smoothing scale is $\Sigma_{\rm smooth} = 15\hmpc$. 

\begin{figure}
    \begin{center}
        \includegraphics[width=\columnwidth]{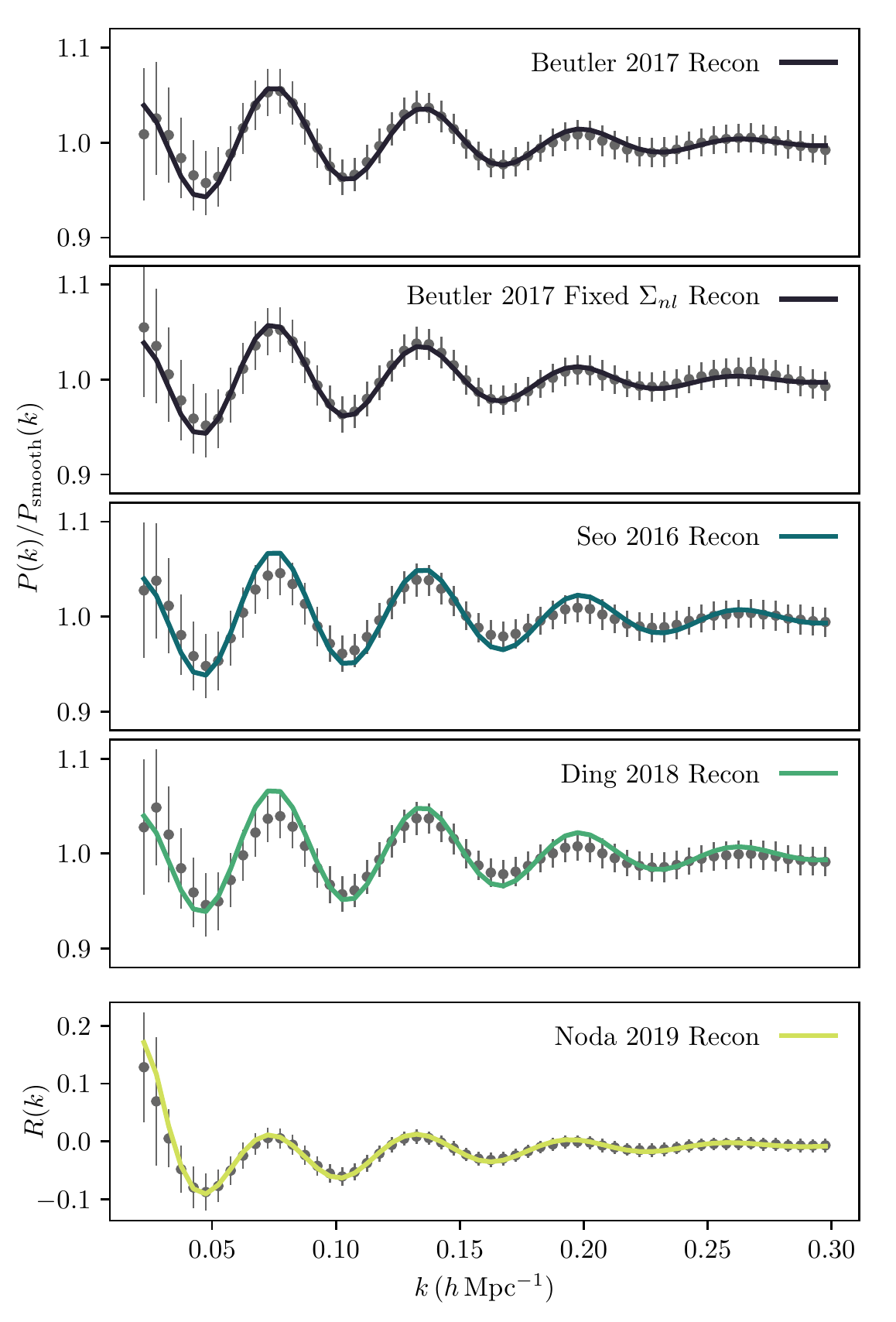}
    \end{center}
    \caption{Average post-reconstruction power-spectrum measurements from mocks based on BOSS-DR12 data in the North Galactic Cap between $0.5<z<0.75$. Each panel shows the measurements and a different maximum \textit{a posteriori} model from \Cref{sec:models} divided by the best-fitting smooth model. The error bars are the errors on a single realisation, \textit{not} the error on the average. For the {\noda} model we show measurements and models after applying the BAO extraction algorithm in \Cref{eq:extractor} as this is what the model was fit to.}
    \label{fig:pk_avg_bestfits}
\end{figure}

\begin{figure}
    \begin{center}
        \includegraphics[width=\columnwidth]{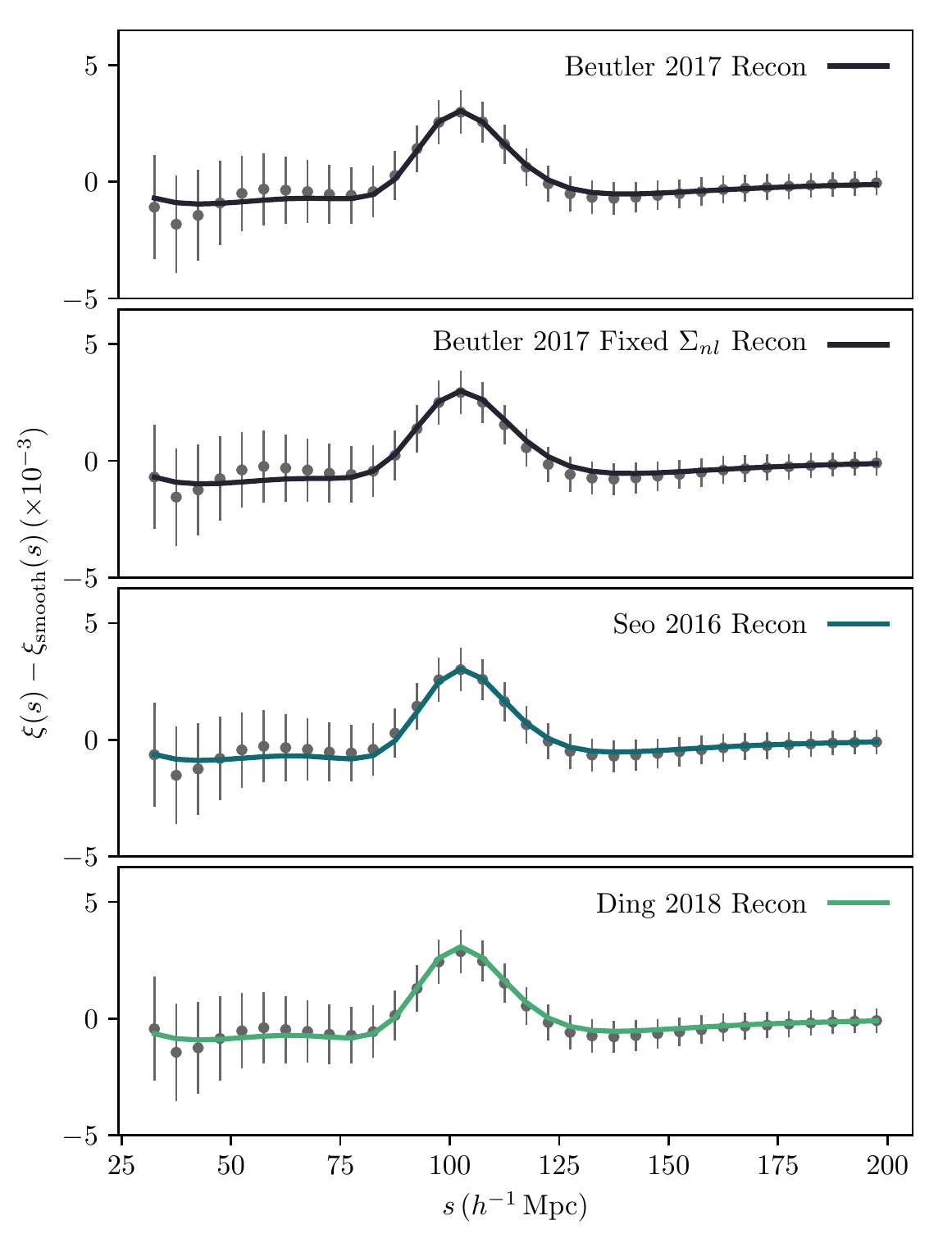}
    \end{center}
    \caption{Average post-reconstruction correlation function measurements from mocks based on BOSS-DR12 data in the North Galactic Cap between $0.5<z<0.75$. Each panel shows the measurements and a different maximum \textit{a posteriori} model from \Cref{sec:models} after subtracting the best-fitting smooth model. The error bars are the errors on a single realisation, \textit{not} the error on the average.}
    \label{fig:xi_avg_bestfits}
\end{figure}

The average post-reconstruction power spectrum measurements from the mocks are shown in \Cref{fig:pk_avg_bestfits} alongside the maximum \textit{a posteriori} model for each of the methods described in \Cref{sec:models}. For each, we have divided out the best-fit smooth model except for the {\noda} model where we plot $R(k)$ given by \Cref{eq:extractor} as this is what the model was fit to. In all cases we see excellent agreement between the measurements and models and although both the {\seo} and {\ding} models do seem to slightly underestimate the BAO damping, our more detailed comparison in \Cref{sec:pk} shows that this does not result in biased constraints. \Cref{fig:xi_avg_bestfits} shows the corresponding measurements and models for the correlation function after the smooth model has been subtracted. Again we see excellent agreement for all models, but a more comprehensive comparison of their performance is given in \Cref{sec:xi}.

We elect to test model differences with only a set of simulations as the input cosmology of the simulations is known, the dataset represents the single most constraining and publicly available simulation suite, and the large number of realisations allows for fitting to an effectively noise free average, as well as individual measurements.


\begin{figure*}
    \begin{center}
        \includegraphics[width=0.9\textwidth]{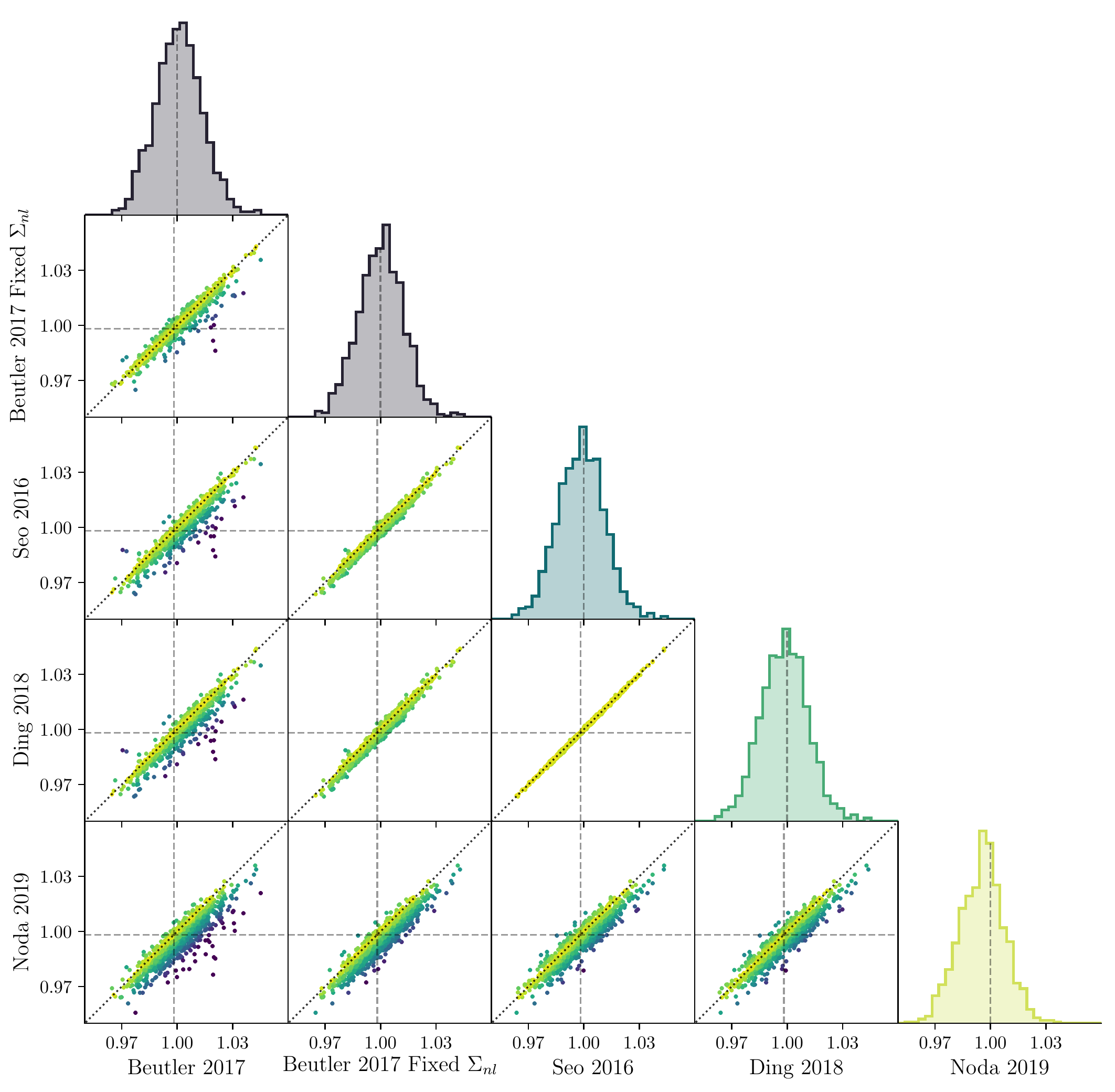}
    \end{center}
    \caption{Direct $\alpha-\alpha$ comparison from fits to each of the 1000 post-reconstruction mocks for all methods. Scatter points are coloured according to $\abs{\alpha_1 - \alpha_2}$. Diagonal sub-plots show histograms of the fits. The similarity of {\seo} and {\ding} manifests in the high correlation between the methods, with both {\beutler} and {\noda} showing increased scatter between models, with the {\noda} method producing the most scatter when comparing to other models, even though its $\alpha$ distribution without reference to other models has similar width. The importance of fixing $\Sigma_{nl}$ in the {\beutler} method can be seen in the reduced scatter and fewer outliers in the fixed model.}
    \label{fig:pk_avg_individual}
\end{figure*}


\section{Power Spectrum Model Validation}
\label{sec:pk}
In verifying the models against the BAO power spectrum signal, we are primarily concerned with recovering an unbiased value for the dilation parameter $\alpha$, as bias on $\alpha$ translates directly into biased cosmology.

\begin{table*}
    \centering
    \caption{Recovery of $\alpha$ when fitting to the mock for each tested model for both pre- and post-reconstruction measurements and for both power spectrum and correlation function fits. Our expected recovery value is $\alpha=0.9982$. We show both fits to the mock-means using the standard covariance (with statistics showing the maximum likelihood and 68\% confidence iso-likelihood regions), and also the mean $\alpha$ recovered by fitting each of the 1000 mocks individually. For this case, the uncertainty shown is the standard deviation of the 1000 $\alpha$ values to help with direct comparison to the mock-mean fit. To get the Monte-Carlo uncertainty on the mean, the uncertainty would be reduced by a factor of $\sqrt{1000}$. The slightly higher $\alpha$ values for the individual mock calculate can be attributed to the skewed nature of the distributions seen in Figures~\ref{fig:pk_avg_individual} and \ref{fig:xi_avg_individual_alphacomp}.}
    \label{tab:model_params}
    \resizebox{\textwidth}{!}{
    \begin{tabular}{l|llll|llll}
        \hline
		Model     & \multicolumn{2}{c}{$P(k)$ Recon $\alpha$}  & \multicolumn{2}{c}{$P(k)$ Pre-recon $\alpha$} & \multicolumn{2}{c}{$\xi(s)$ Recon $\alpha$}  & \multicolumn{2}{c}{$\xi(s)$ Pre-recon $\alpha$} \\ 
		          & Mock-Mean  & Ind. Mocks  & Mock-Mean  & Ind. Mocks & Mock-Mean  & Ind. Mocks  &  Mock-Mean  & Ind. Mocks  \\ 
		\hline
		{\beutler} & $0.999^{+0.013}_{-0.011}$ & $1.001 \pm 0.013$ & $1.003^{+0.021}_{-0.020}$  & $1.006 \pm 0.021$ &$0.999^{+0.016}_{-0.013}$  & $1.002 \pm 0.014$ & $1.001^{+0.024}_{-0.021}$ & $1.005 \pm 0.021$\\ 
		{\beutler} Fixed $\Sigma_{nl}$ & $0.999^{+0.013}_{-0.011}$ & $1.000 \pm 0.012$ & $1.003\pm 0.019$ & $1.004 \pm 0.020$ & $1.000^{+0.015}_{-0.013}$  & $1.001 \pm 0.013$ & $1.002^{+0.022}_{-0.020}$ & $1.003 \pm 0.020$\\ 
		{\seo} & $0.999\pm 0.010$ & $0.999 \pm 0.013$ &$0.995^{+0.016}_{-0.012}$  & $0.996 \pm 0.020$ & $1.000^{+0.014}_{-0.013}$ & $1.001 \pm 0.013$ & $1.000^{+0.021}_{-0.019}$ &  $1.003 \pm 0.020$\\ 
		{\ding} & $0.998^{+0.010}_{-0.009}$ &  $0.998 \pm 0.013$ &$0.997^{+0.015}_{-0.014}$  &  $0.998 \pm 0.020$ &$1.000^{+0.013}_{-0.014}$  & $1.000 \pm 0.013$ & $1.000^{+0.021}_{-0.020}$ & $1.003 \pm 0.020$\\ 
		{\noda} & $0.995^{+0.012}_{-0.010}$  & $0.995 \pm 0.012$ & $1.012^{+0.017}_{-0.016}$ & $1.014 \pm 0.017$ & --- & --- & --- & --- \\ 
		\hline
    \end{tabular}
    }
\end{table*}

\subsection{Validating $\alpha$ recovery}

In this section we will validate $\alpha$ recovery in two separate ways. First we fit to the mean of our mock data --- representing fits to a single, effectively noise-free, sample of incredible statistical power. We do not modify the covariance calculated from the mocks when fitting the mean (although we could reduce it by a factor of $\sqrt{1000}$ to represent the increased statistical power), and so the uncertainty on the mock-mean fits is still representative of an individual realisation, even though we would expect our parameter inferences to recover the truth values to greater accuracy. We also fit to each of the 1000 mocks available independently, sacrificing computational efficiency for the ability to investigate the $\alpha-\alpha$ correlations between the different models (which can be seen in \Cref{fig:pk_avg_individual}).

For the {\beutler} model, we test both a model with free $\Sigma_{nl}$ and also the model with $\Sigma_{nl}$ fixed to the value from a global fit to the mean-mock data. We note here that we were required to do a global nested sampling fit to recover $\Sigma_{nl}$ with confidence, as optimisation routines frequently gave different answers depending on their starting location in parameter space. We present both models here to illustrate the effect fixing $\Sigma_{nl}$ has on outliers when fitting the individual mocks. We find that for post-reconstruction measurements, $\Sigma_{nl} = 6.0\,h^{-1}\mathrm{Mpc}$ and for pre-reconstruction, the best fit $\Sigma_{nl} = 9.3\,h^{-1}\mathrm{Mpc}$.

\begin{figure}
    \begin{center}
        \includegraphics[width=\columnwidth]{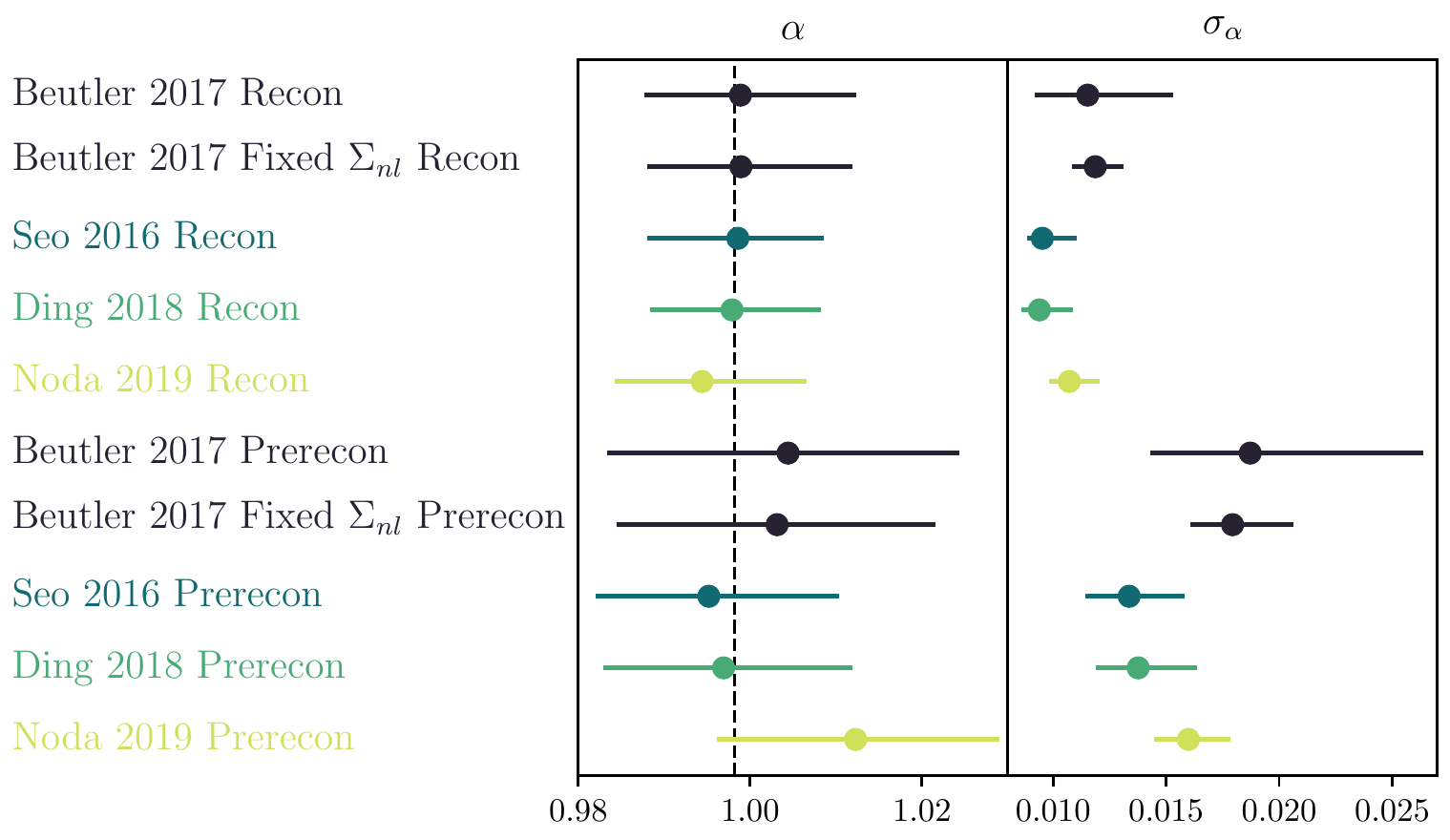}
    \end{center}
    \caption{The first column shows the 68\% confidence interval summary of $\alpha$ when fitting to the mean mock data using the power spectrum method. The second column shows 68\% confidence interval on the distribution of errors when fitting $\alpha$ for each of the thousand individual mocks. Fixing $\Sigma_{nl}$ to the best fit value does not help reduce the uncertainty on the recovered $\alpha$ value when fitting the mean, however does significantly reduce the spread in the errors on $\alpha$ as shown in the second column. {\seo} and {\ding}, being highly similar methods, give comparable uncertainty and recovery of $\alpha$. The {\noda} model displays bias in $\alpha$, potential causes for this are discussed in Sec.~\ref{sec:noda}. }
    \label{fig:pk_avg_summary2}
\end{figure}

\Cref{fig:pk_avg_summary2} shows the fits to the mock-mean data. A tabular version of this figure, combined with the correlation function results, is presented in \Cref{tab:model_params}. All methods barring {\noda} pre-reconstruction recover the expected $\alpha=0.9882$ with bias less than half of the statistical error. The {\noda} method run on pre-reconstruction data recovers $1.012^{+0.017}_{-0.016}$. As our mock-mean is computed by the average of 1000 mocks, this represents a $\sim27\sigma$ detection of a $\sim 0.86\sigma$ bias (given the statistical power of an individual mock realisation). We attribute this to the inability of SPT to accurately model the non-linear power spectrum, which we investigate further in \Cref{sec:noda_nonlinear} using an alternative model for these components. Curiously, this was not found in the original {\noda} analysis. In addition, contrary to their work, we do not find significantly tighter $\alpha$ constraints after applying the BAO extraction procedure. More detailed tests on the claims found in {\noda} are presented in Section.~\ref{sec:noda}.

\Cref{fig:pk_avg_individual} continues our investigation into the $\alpha$ recovery of our power spectrum models, showing each individual post-reconstruction mock's mean recovered $\alpha$ against all other models. Comparing the {\beutler} models with and without fixing $\Sigma_{nl}$ shows that not fixing the value significantly increases the deviation from {\beutler} to the other models and creates outliers clearly visible in the scatter plots in the first column. Additionally, we can also see the extreme correlation between the {\seo} and {\ding} models, which may be expected due to the fact the models only differ in their propagators; which although theoretically quite different, give very similar BAO shapes. Finally, we also see larger scatter between the {\seo} and the {\noda} models, which is not surprising given that the {\noda} model represents both a different model of the power spectrum and the application of the BAO Extractor, which may indicate there is some benefit in combining these results. As a further investigation we apply the BAO extraction procedure to different models in Section~\ref{sec:noda} and the impact of combining measurements using different methods in \Cref{sec:consensus}.

With the exception of the bias in the {\noda} results and larger scatter in general arising from less precise constraints, fits to the pre-reconstruction mocks give qualitatively similar results to \Cref{fig:pk_avg_individual} and are thus not explicitly illustrated. Instead we turn to look at the errors on $\alpha$ recovered from each method.

\begin{figure*}
    \begin{center}
        \includegraphics[width=0.85\textwidth]{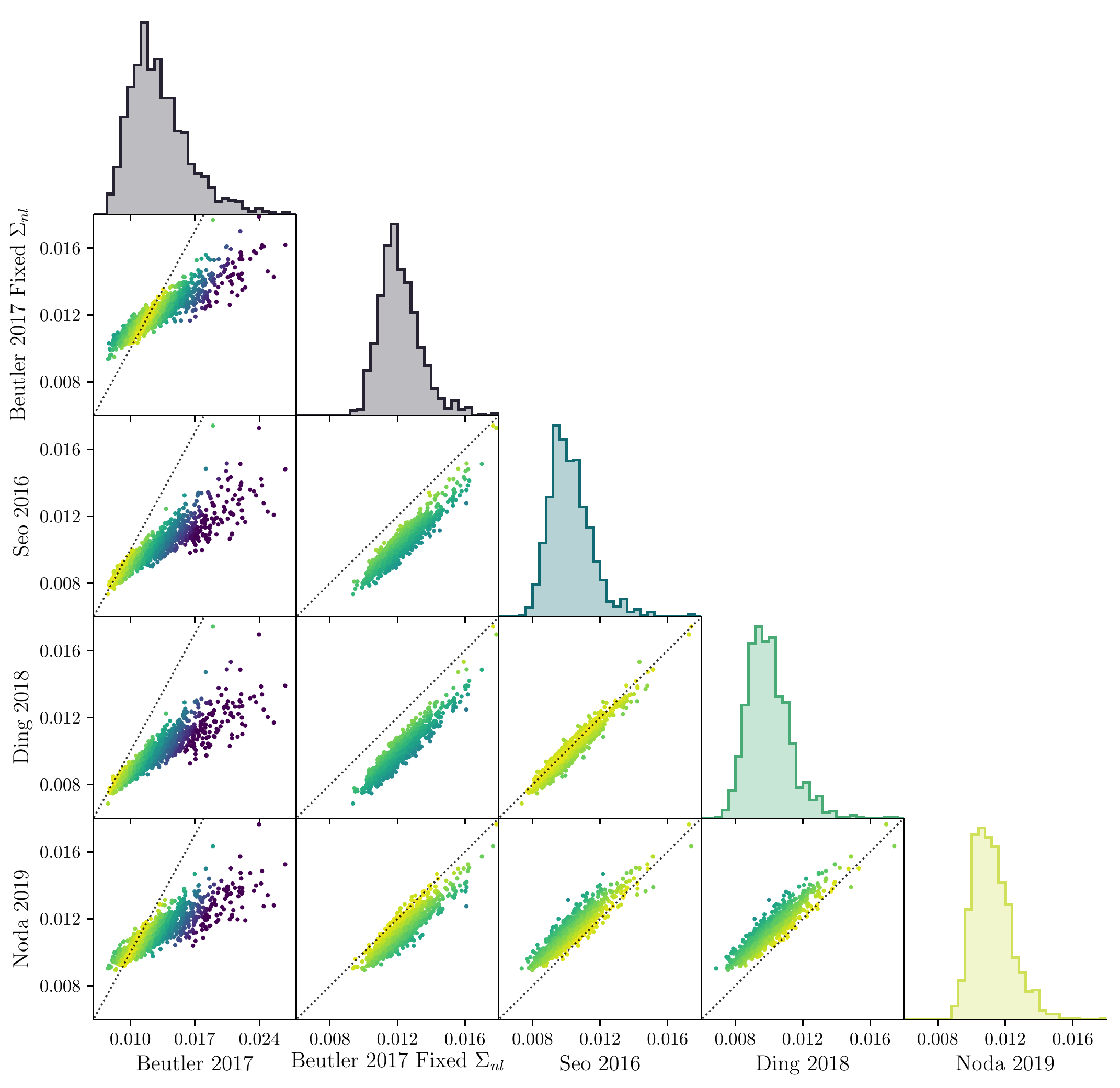}
    \end{center}
    \caption{A comparison of the standard deviation in $\alpha$ from fits to each of the 1000 post-reconstruction mocks for all methods. Scatter points are coloured according to $\abs{\sigma_{\alpha_1} - \sigma_{\alpha_2}}$. Allowing the BAO damping parameters to vary, as in the {\beutler} model, significantly increases the typical uncertainty compared to other methods. Again the similarity of {\seo} and {\ding} manifests in the high correlation in the errors although the {\ding} model returns the tightest constraints on average post-reconstruction. Results for pre-reconstruction are qualitatively similar, although in this case the {\seo} model does marginally better.}
    \label{fig:pk_err_individual}
\end{figure*}

\subsection{$\alpha$ Errors}

Using our fits to the individual mocks we find a significant change in the typical errors on $\alpha$ recovered using each of the different methods. This is shown in \Cref{fig:pk_err_individual} where we plot the recovered errors for the models. The $16^{th}$, $50^{th}$ and $84^{th}$ percentiles of the distribution of mock errors for each model are also shown in \Cref{fig:pk_avg_summary2} in the second column, and show the variance of the model uncertainty across the thousand mocks. Generally, the precision with which $\alpha$ is recovered scales with the strength of the prior on the BAO damping although the {\seo} and {\ding} models perform noticeably better than simply fixing $\Sigma_{nl}$ within the {\beutler} model. In what follows we attribute this mostly to these models under-reporting the errors on some of the mocks. More specific findings can be summarised as follows:
\begin{itemize}
    \item {Although allowing $\Sigma_{nl}$ to vary in the {\beutler} model generally causes larger uncertainty in the recovered $\alpha$ value, it actually improves the constraints on $\alpha$ in $\sim24\%$ and $\sim36\%$ of mocks pre- and post-reconstruction respectively. This is due to noise altering the shape of the BAO feature and making it more prominent, and a strong correlation between the uncertainty on $\alpha$ and the best fit value of $\Sigma_{nl}$. As recently highlighted in \cite{Ruggeri2019}, it may actually be undesirable for a smaller error to be reported in this case, as it is not necessarily a result of the underlying physics or cosmology.}
    \item {The errors for the {\seo} and {\ding} models are highly correlated, as expected given the high degree of correlation between their best-fit alpha values. The {\ding} model does perform slightly better post-reconstruction, whilst {\seo} does better pre-reconstruction. However, as shown later in \Cref{fig:chi2}, the constraints on $\alpha$ from both of these models across all the mock realisations are not $\chi^{2}$ distributed about the expected value of $\alpha=0.9882$, leading us to conclude that these models are under-reporting the error on a mock-by-mock basis. This is discussed further in \Cref{sec:consensus}, where we also provide a method to remedy this/choose between the various models in this study.}
    \item{From \Cref{fig:pk_avg_summary2}, we can see that the {\noda} model rarely performs best, although the comparison is a little unfair as this model returns much more representative errors than {\seo} or {\ding}. Nonetheless, our findings are again in contrast to the claimed improvement offered by the BAO extractor in \cite{Noda2019}, a point we investigate further later. However, those mocks where it does do better seem to be those with larger errors, leading us to conclude that the BAO extractor may be better at identifying and fitting the BAO peak when the data is noisy.}
    \item{We found no correlation between the uncertainty for any of the methods and the recovered value of alpha.}
\end{itemize}

\subsection{Numerical Efficiency}

Our current tests have shown a small difference between the Fixed {\beutler}, and the {\seo}, {\ding} and {\noda} methods at recovering cosmological parameters with current-generation survey statistical power. However, there does exist significant discrepancy in their numerical cost that is worth discussion. Current power spectrum implementations --- including invocation of CAMB to determine the linear power spectrum at a given cosmology --- have the {\beutler} method computing a likelihood in under a millisecond. The {\seo} and {\ding} methods, with their more complicated propagator and integral over $\mu$, require approximately 40ms to compute, giving the {\beutler} method an order of magnitude benefit in computational efficiency. For surveys that might want to engage in systematic tests of model parametrisations, parameter priors or other tests, an increase in computational efficiency can directly tie to an increased number of tests, or increase statistical power of those tests.

With that in mind, it must be highlighted that the computational benefit {\beutler} provides is due to the fact we are fitting an isotropic model in this work. An anisotropic model, which is of greater interest to future surveys and analyses, will have the $\mu$ integration term in all models, naively rendering all models approximately equal in computational efficiency, provided there are no significant extra computational costs involved in determining the terms of the integral. However, if the models are symmetric about $\mu$ it may be possible these can be solved analytically, or separated out into terms each of which can be treated in the same way as the isotropic {\beutler} model. This would preserve the numerical performance but would be highly dependent on the mathematical formulation of the particular model.

Of greater importance to run time is the method of sampling our likelihoods. In our investigations, we have compared the Metropolis-Hastings algorithm (MH; \citealt{Hastings1970}) with \textsc{emcee} \citep{emcee} and the nested sampling algorithm found in \textsc{dynesty} \citep{dynesty}. We find for all algorithms the likelihood surface is simpler (a closer approximation to a uni-modal Gaussian) for fitting the correlation function models. The slowest sampling method implemented in \textsc{Barry}, \textsc{dynesty}, consistently gave reliable samples of the likelihood with minimal tuning of the initial algorithm conditions. Using, for instance, \textsc{emcee}, we found that greater care needed to be taken to ensure convergence and adequate burn-in for each individual mock. As such, \textsc{dynesty} was adopted for all the model fitting in this paper.

\begin{figure}
    \begin{center}
        \includegraphics[width=\columnwidth]{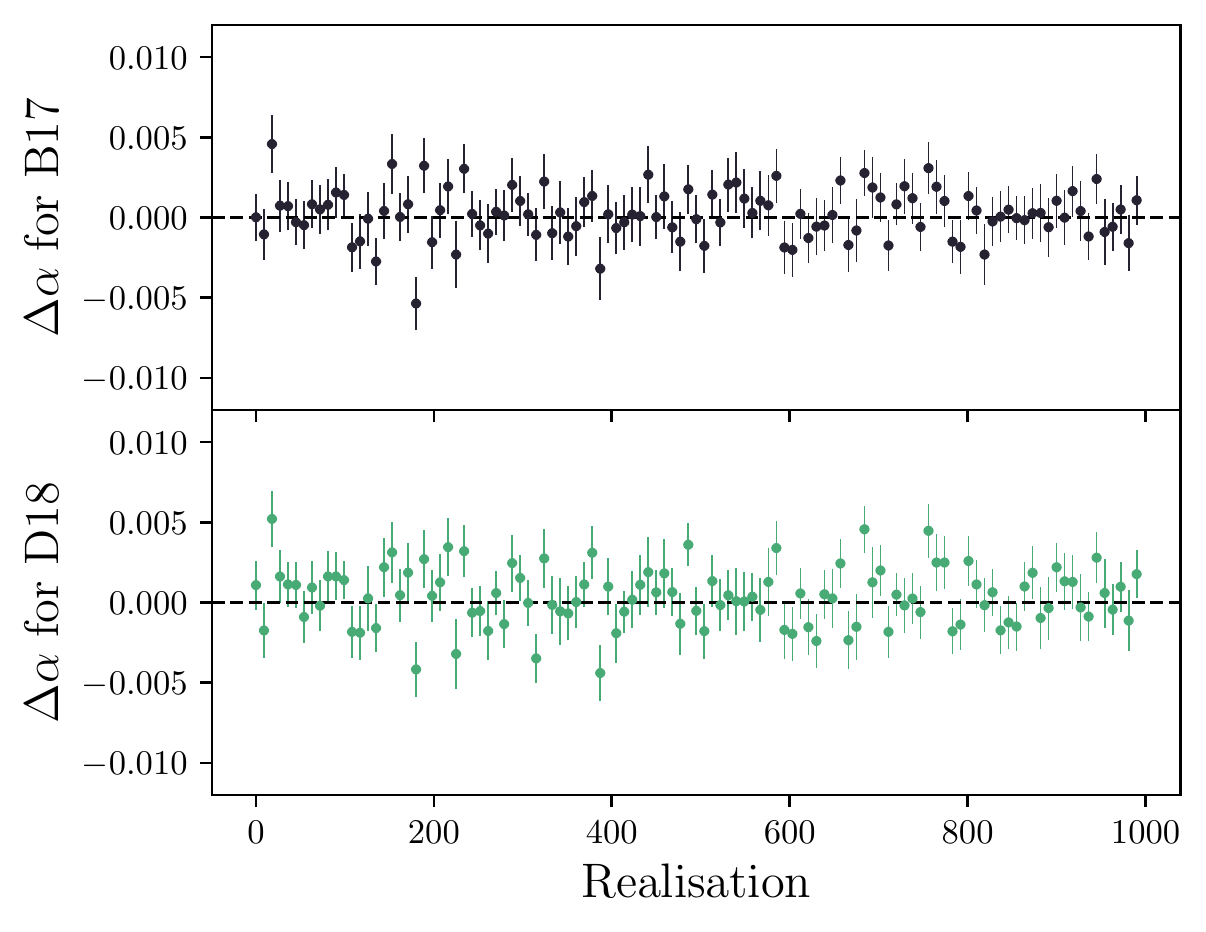}
    \end{center}
    \caption{ Comparing {\beutler} and {\ding} methods with and without the BAO extractor applied. The top row shows the difference in mean $\alpha$ between models with and without the extractor applied for the {\beutler} model. The bottom row shows {\ding}. Error bars represent the variance of both model's uncertainty on $\alpha$ with covariance estimated using the correlation in $\alpha$ values for all 1000 mocks. Every ninth realisation is shown for clarity. The scatter between the models with and without the extractor is approximately $0.002$ for both models and is consistent with the reported uncertainty on $\alpha$. This shows that the presence of the extractor is not introducing scatter or bias that is seen in the results from {\noda} in Fig.~\ref{fig:pk_avg_individual}. }
    \label{fig:ding_baoextractor_alphacomp}
\end{figure}

\subsection{Noda}
\label{sec:noda}

A novel method of extracting the BAO feature has been presented in \citet{Noda2017, Nishimichi2018} and has been applied onto the SDSS DR12 dataset in \citet{Noda2019}. The extractor is designed to separate out the long and short-scale effects in the evolution of the BAO feature, allowing for for the extraction of the oscillating component of the BAO feature and thus requiring a reduced number of nuisance parameters. Chief among the benefits of the {\noda} method is the claim of increased accuracy by an approximate factor of two over traditional analyses.

Our initial evaluation of the {\noda} method did not support these claims, and was also unable to recover unbiased $\alpha$ values for the pre-reconstruction data, as shown in \Cref{sec:pk}. We investigate this further here.

\subsubsection{The effect of the extractor}

First, we wish to determine the effects of the extraction algorithm itself, independent of the {\noda} theoretical model. We note that although the {\noda} model is `bundled-in' with the extraction procedure in the references for this method, there is no reason the extractor can not be applied to the other models tested in this work. This is what we do here, and implement the extractor method on both the {\beutler} and {\ding} models, comparing the outputs from fitting all thousand mocks for those two methods, both with and without the extractor applied. Results are shown in \Cref{fig:ding_baoextractor_alphacomp} and laid out explicitly in \Cref{tab:extractor_effect}. These results reveal that the extractor procedure is not causing any bias in the results, but is also not improving the alpha measurements on average.

We also investigate the effect of the anchor point chosen by {\noda}. As the {\noda} method attempts to constrain $\alpha$ from the extracted signal, and constrain $A$ (see eqs. \ref{eq:noda1} and \ref{eq:noda2}) from the raw power spectrum, a composite fit to both is performed in {\noda} by designating specific bins to either have the extractor applied or not. The method used to split the data from {\noda} is to define both a lower limit $k_1$, and two additional higher-$k$ anchor points, $k_2$ and $k_3$. Below $k_1$, they utilise the power spectrum. They then add onto these bins (there are three with $k < k_{1}$ in the case of {\noda}) two additional bins, giving five data points of raw power spectrum, with the rest using the extracted power spectrum. Expressed in a logical statement, bins that were assigned to the power spectrum instead of the extracted signal satisfy $(k < k_1 \lor k = k_2 \lor k = k_3)$. We vary $k_3$ in one test ($k_3 \approx 0.1775 \hmpc$ for {\noda}) from $0.13 \hmpc$ to $0.24 \hmpc$ in increments of $0.01$ and found that the choice of $k_3$ induces scatter in $\alpha$ with standard deviation $0.006$ (and scatter of $0.6$ in $A$), making it a potential source of systematic uncertainty in future analyses, but not a significant contribution in this investigation. A separate test that varied $k_1$ found negligible scatter for viable lower bounds - $0.02 \hmpc$ to $0.07 \hmpc$.

\begin{table}
	\centering
	\caption{Determining the effect of the BAO Extractor on the {\ding} and {\beutler} models. Note that, due to the index mixing, both models had to keep their polynomial terms, resulting in the BAO Extractor method not reducing model dimensionality.}
	\label{tab:extractor_effect}
	\begin{tabular}{lccr} 
		\hline
		Model & $\langle \alpha \rangle$ & $S_\alpha$ & $\langle \sigma \rangle$ \\
		\hline
		{\ding}                & 0.998 & 0.013 & 0.010\\
		{\ding} + Extractor    & 0.999 & 0.013 & 0.010\\
		{\beutler}             & 1.000 & 0.012 & 0.012\\
		{\beutler} + Extractor & 1.000 & 0.013 & 0.012\\
		\hline
	\end{tabular}
\end{table}

\subsubsection{On the choice of non-linear correction}
\label{sec:noda_nonlinear}

\begin{figure}
    \begin{center}
        \includegraphics[width=\columnwidth]{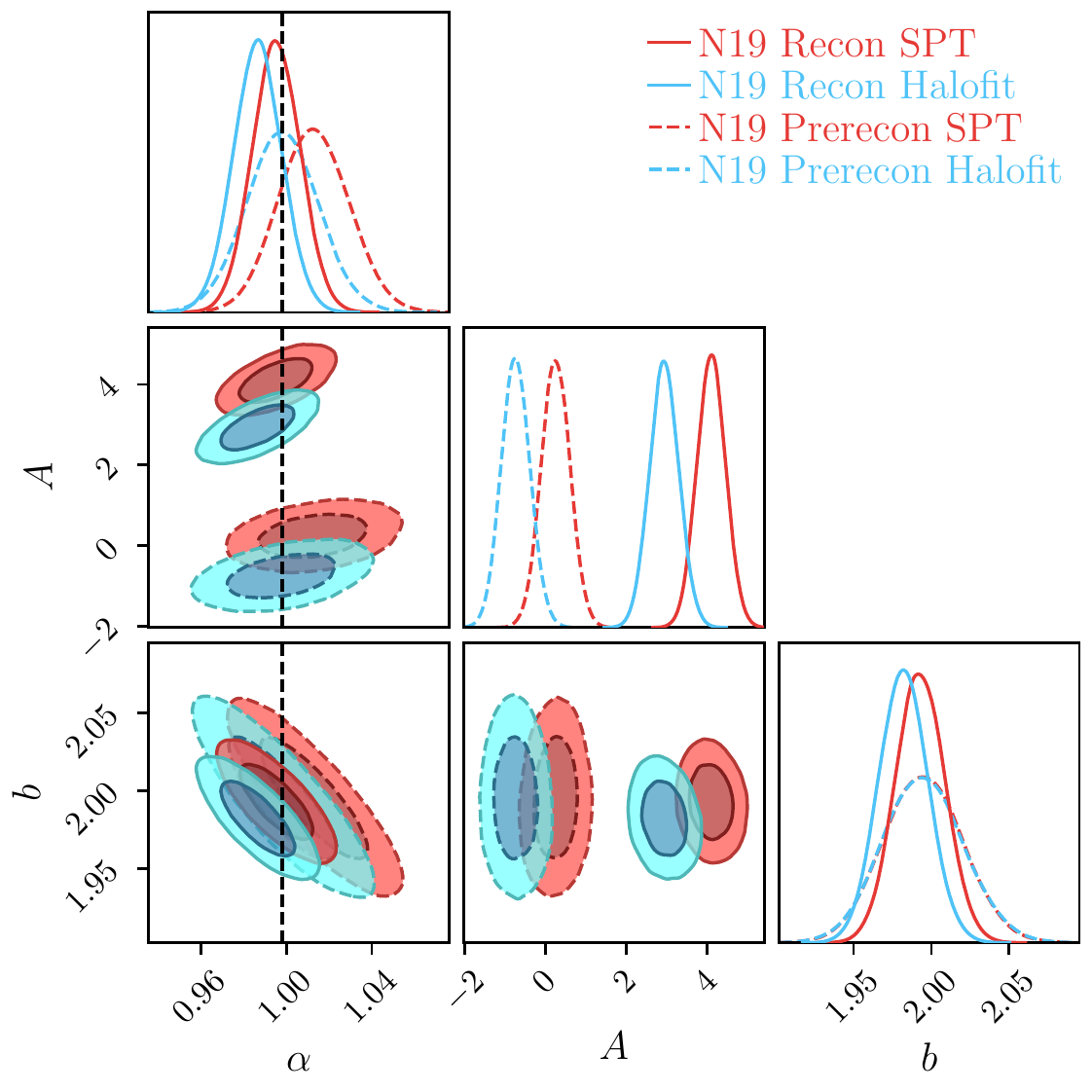}
    \end{center}
    \caption{Comparing the results of the {\noda} when changing the non-linear correction in the model between SPT and `Halofit'. SPT (the closest to the original model from {\noda}) performs better post-reconstruction as reconstruction makes the model more linear, whilst `Halofit' performs better on pre-reconstruction data and is able to produce unbiased results.}
    \label{fig:noda_nl}
\end{figure}

The {\noda} model implementation utilises non-linear terms for the smooth power spectrum derived from standard perturbation theory. Without the extra parameters found in other models that increase model flexibility, the results of the {\noda} model will be more dependent on the choice of underlying non-linear correction. To test if the choice of correction is a potential source of bias (and indeed \textit{the} source of bias we found in the pre-reconstruction fits in \Cref{fig:pk_avg_summary2}), we have implemented both the standard SPT-based non-linear terms and non-linear `halofit' correction \citep{Takahashi2012, Mead2016} paired with the velocity divergence/cross-power spectrum fitting formulae from \cite{Jennings2012b}. As shown in \Cref{fig:noda_nl}, changing corrections induces a shift in $\alpha$. We find that for post-reconstruction measurements, the SPT model performs better, as would be expected due to reconstruction's linearising effect. For pre-reconstruction, we find that the `halofit' correction performs better, giving an unbiased estimate of $\alpha$. We conclude that the {\noda} method is only biased due to the choice of non-linear correction, and that adopting different prescriptions pre- and post-reconstruction as done here (and as also done for e.g., the {\seo} model) or just generally using more accurate non-linear models (as in \citealt{Nishimichi2018}) can resolve this.

\subsubsection{On fixing model parameters}

We next examine fixed parameters in the {\noda} method.  Specifically, {\noda} sets the parameter $\gamma_{rec}$ --- which models the reduction in BAO damping from reconstruction --- as $4$ for post-reconstruction, and $1$ for pre-reconstruction, using the approximate halving in dampening scale reported from theoretical predictions in \citet{Eisenstein2007}. However, this is only approximately correct, and so we also investigate the effect of freeing the $\gamma_{rec}$ parameter in our fits. Additionally, {\noda} also fix the bias factor $b$ to a preliminary initial fit of the power spectrum monopole and quadrupole up to $k\simeq0.15\hmpc$, stating that the value fit for $b$ has little impact on the final result. Our initial fits with the {\noda} model shown previously \textit{do not} fix $b$. We thus test the effect of fixing $b$ using an initial fit only with the monopole up to the same $k$ range as {\noda}. Additionally, {\noda} keeps the growth rate parameter $f$ fixed, but this is common across many BAO models, and so we keep this fixed as a baseline.

\begin{figure}
    \begin{center}
        \includegraphics[width=\columnwidth]{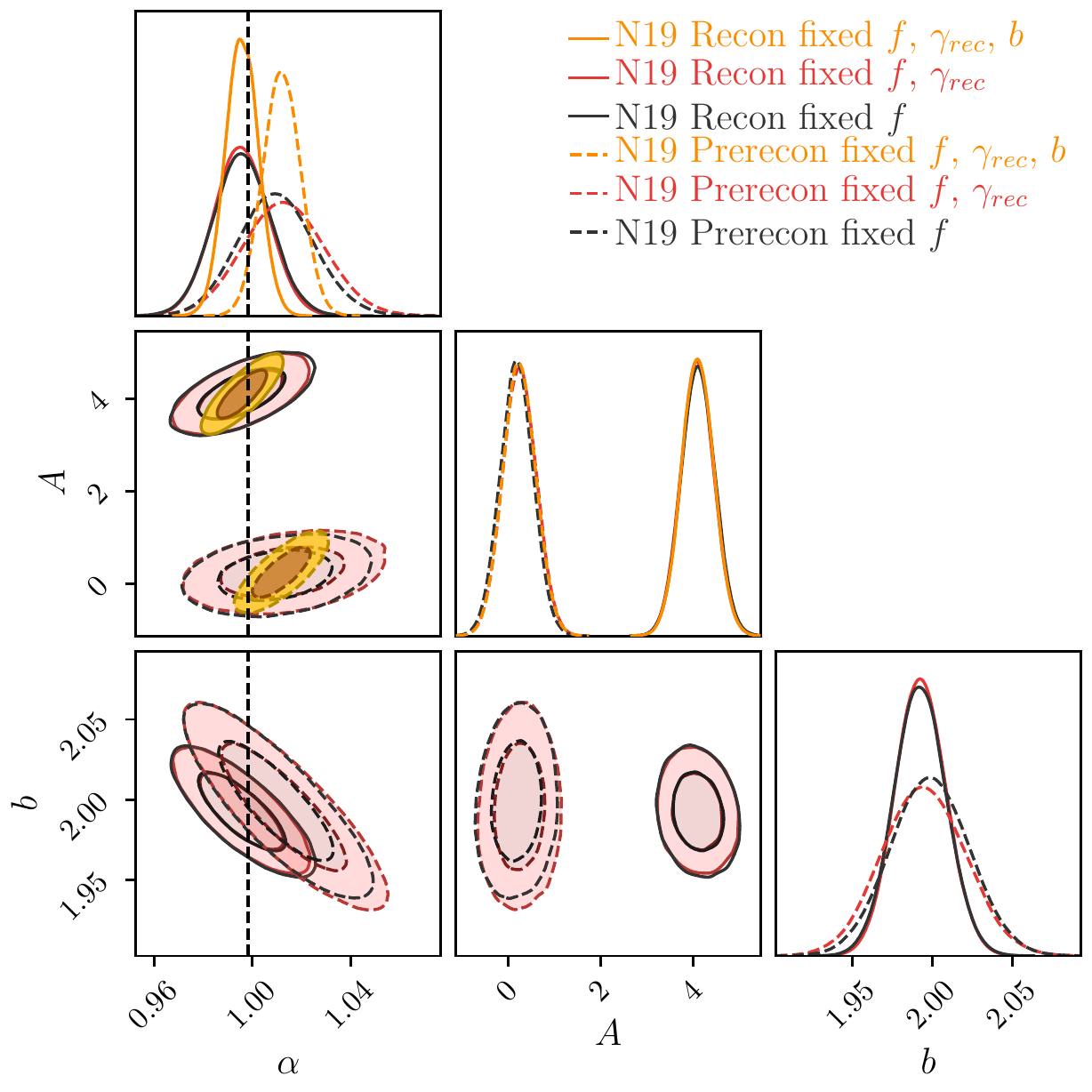}
    \end{center}
    \caption{Parameter constraints when model parameters are freed or fixed. The majority of the tighter constraint on $\alpha$ reported in {\noda} appears due to the fixing of $b$, and allowing $b$ to be free produces a significant increase in uncertainty. Furthermore, the value to which $b$ is fixed has implications on $\alpha$, with higher values for $b$ preferring lower fits to $\alpha$. Fitting $b$ initially using a different range of $P(k)$ causes shifts of up to $0.02$ in $b$, inducing similar shifts in $\alpha$. Shifting the fixed value of $\gamma_{rec}$ also some effect on $\alpha$, with higher $\gamma_{rec}$ preferring lower $\alpha$. However, compared to the statistical error, the effect from $\gamma_{rec}$ is not significant.}
    \label{fig:noda_avg}
\end{figure}

The results from these model comparisons are shown in \Cref{fig:noda_avg}. We found that fixing $\gamma_{rec}$ does not significantly modify likelihood surfaces (a higher $\gamma_{rec}$ does prefer a negligibly lower $\alpha$, not shown in \Cref{fig:noda_avg}). However, the primary result is that fixing $b$ has potential to both bias $\alpha$ and is the dominant source of {\noda}'s claim of improved constraints. The post-reconstruction uncertainty on $\alpha$ changes from $\pm0.0066$ to $\pm0.011$ when $b$ is unfixed, bringing the uncertainty inline to that of the other tested models (shown in \Cref{tab:model_params}). Additionally, there is a strong degeneracy between $b$ and $\alpha$ (correlation value $-0.81$), and choice in how to fit $b$ initially (for example, modifying the $k$ range) can result in a difference of up to $0.02$ in the best fit value, which shifts $\alpha$ by a similar amount. These findings contradict the claims of {\noda} that the fixing of $b$ has little impact on the final result.

In this section, we have evaluated our implementation of the {\noda} model. We find that the claims of improved fitting constraints are primarily the result of fixing model parameters to best-fit values, and not the result of the new BAO Extractor technique. Additionally, we recommend a more thorough investigation in how to utilise index mixing (the choice of $k_1$, $k_2$, $k_3$) to be able to fit the combined power spectrum and extracted power spectrum without introducing potential bias.

 

\section{Correlation Function Model Validation}
\label{sec:xi}

In addition to testing our implemented models against the power spectrum, we also perform fits against the correlation function. This requires minimal changes in the models; removing explicit treatment of window function effects in the likelihood, implementing the spherical Hankel transformation into real-space, and a change in the function polynomial shape terms to apply after the transformation. Here we note too that the {\beutler} method in configuration-space is essentially the method presented in \citet{Ross2015}, however for consistency we will continue to reference it as the {\beutler} method for easier comparison. As the complete {\noda} method is specific to power spectrum analysis, it would be an unfair comparison to convert it into a correlation function and thus we leave it out of this section.

For our spherical Hankel transformation we utilise high-resolution Gaussian-damped numerical integration instead of a Fourier transformation, for numerical reasons. Discussion on transformation methods and their numerical stability can be found in \citet[Appendix B]{Hinton2016}.

\subsection{Validating $\alpha$ recovery}

\begin{figure}
    \begin{center}
        \includegraphics[width=\columnwidth]{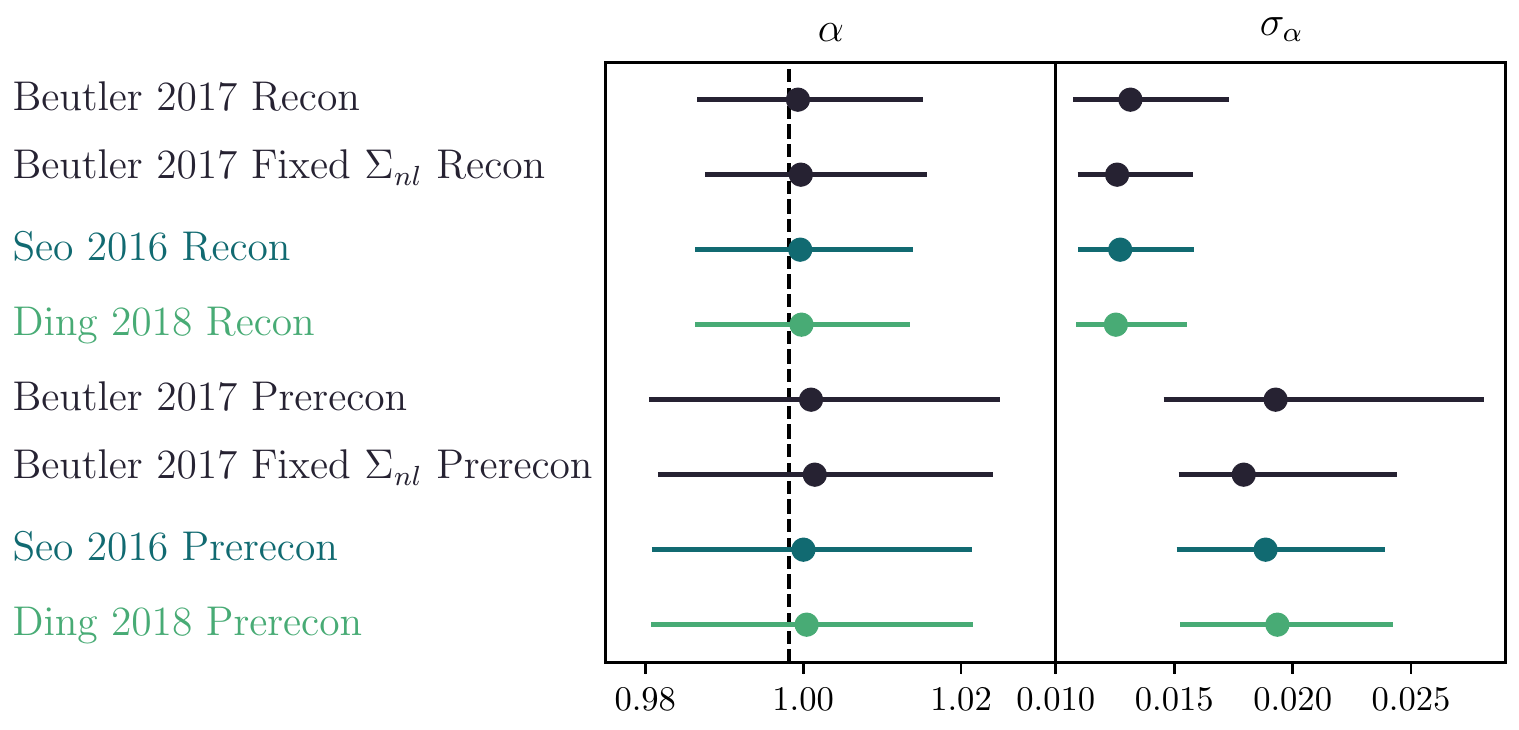}
    \end{center}
    \caption{The first column shows fitting $\alpha$ to the mean $\xi(s)$ of all mocks, using the correlation function method. The second columns shows the distribution of errors on fitting $\alpha$ for each of the thousand individual mocks. Uncertainty is fractionally larger ($\sim 15\%)$ across all methods when compared to fitting the mean $P(k)$, which is shown to not to be an artifact of the different fitting ranges in $k$ and $s$. }
    \label{fig:xi_avg_summary2}
\end{figure}

\begin{figure*}
    \begin{center}
        \includegraphics[width=0.9\textwidth]{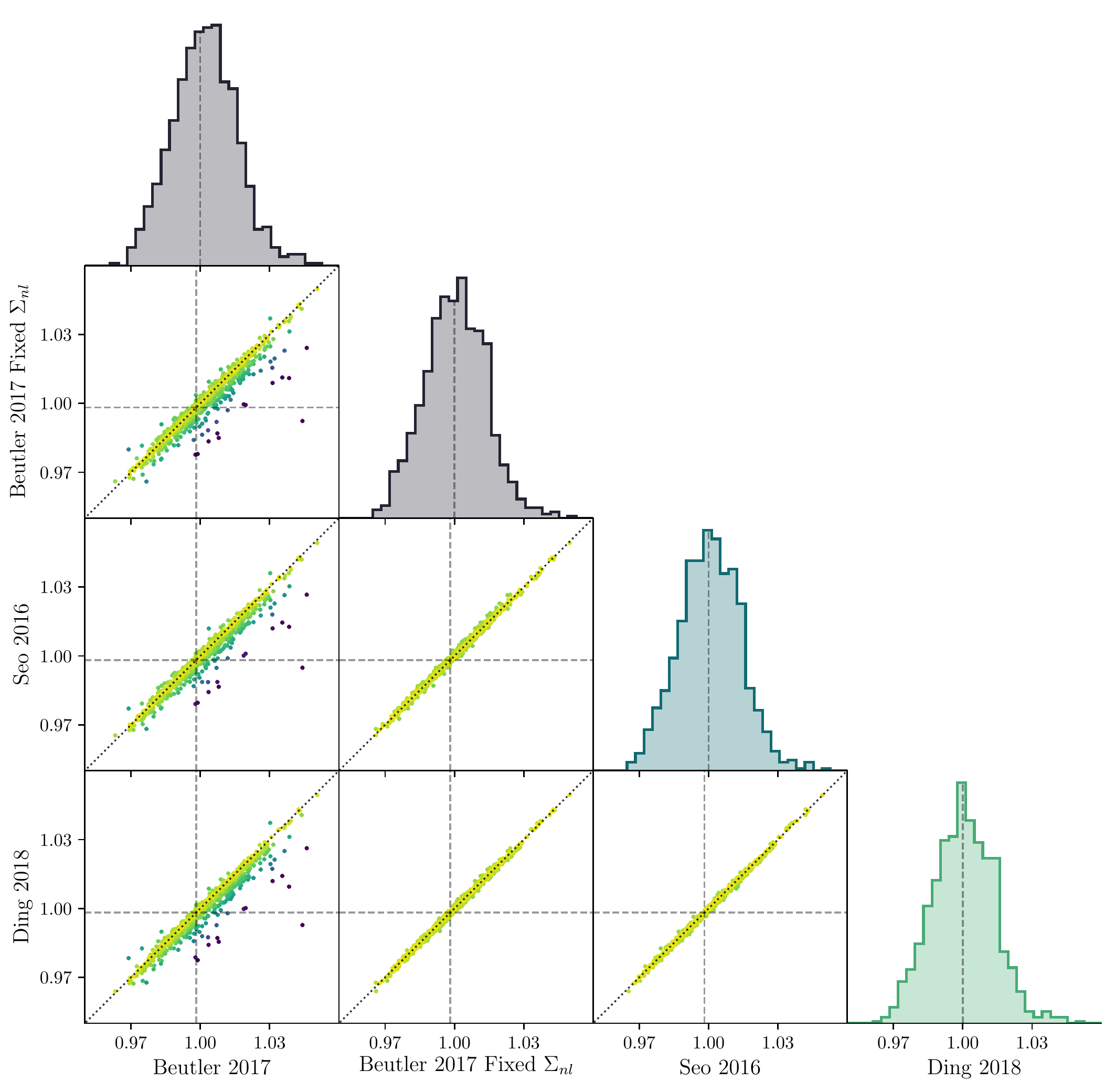}
    \end{center}
    \caption{Direct mean $\alpha-\alpha$ comparison to fits of each of the 1000 post-reconstruction mocks for all correlation function models. Scatter points are coloured according to $\abs{\alpha_1 - \alpha_2}$. Compared to Fig~.\ref{fig:pk_vs_xi_individual_alphacomp}, we see that the {\beutler} fixed $\Sigma_{nl}$ method has correlation on par with that of {\seo} and {\ding}. The importance of fixing $\Sigma_{nl}$ in the {\beutler} method is again highlighted with the presence of outliers in the mock fits when $\Sigma_{nl}$ is kept free.}
    \label{fig:xi_avg_individual_alphacomp}
\end{figure*}

As per the power spectrum section, we fit our models to the mean of all $\xi(s)$ mocks (the mock-mean) and present summaries of our fits in both Fig.~\ref{fig:xi_avg_summary2} and Table~\ref{tab:model_params}. Fits to the individual mocks can be found in Fig.~\ref{fig:xi_avg_individual_alphacomp}. The constraints attained on $\alpha$ when fitting to the mock-mean are less precise than those attained from the power spectrum fits. We investigate this as a potential outcome from our different fitting ranges. Our $k$-range when fitting the power spectrum is $0.02<k<0.30 \hmpc$, which corresponds in real space to approximately $21 < s < 314 \mpch$. However, even when expanding the range in $s$ to agree with the $k$ range, or when reducing the $k$ range to agree with the default $s$ range, we find the same results --- the correlation function models report larger uncertainties than the power spectrum models. This is discussed further in \Cref{sec:consensus}.

Our results show that all models recover unbiased $\alpha$ when fitting to the mock-mean, and an interesting result from Fig.~\ref{fig:xi_avg_individual_alphacomp} is that the correlation between models (specifically the {\beutler} fixed $\Sigma_{nl}$ model) is higher for the correlation function fits than the power spectrum fits. This may suggest that the difference between the models is less significant in the correlation function formulation. The problem of outliers for the {\beutler} model without fixed $\Sigma_{nl}$ is, however, just as prevalent in the correlation function fits as it was in the power spectrum fits.

\subsection{$\alpha$ Errors}
\begin{figure}
    \begin{center}
        \includegraphics[width=\columnwidth]{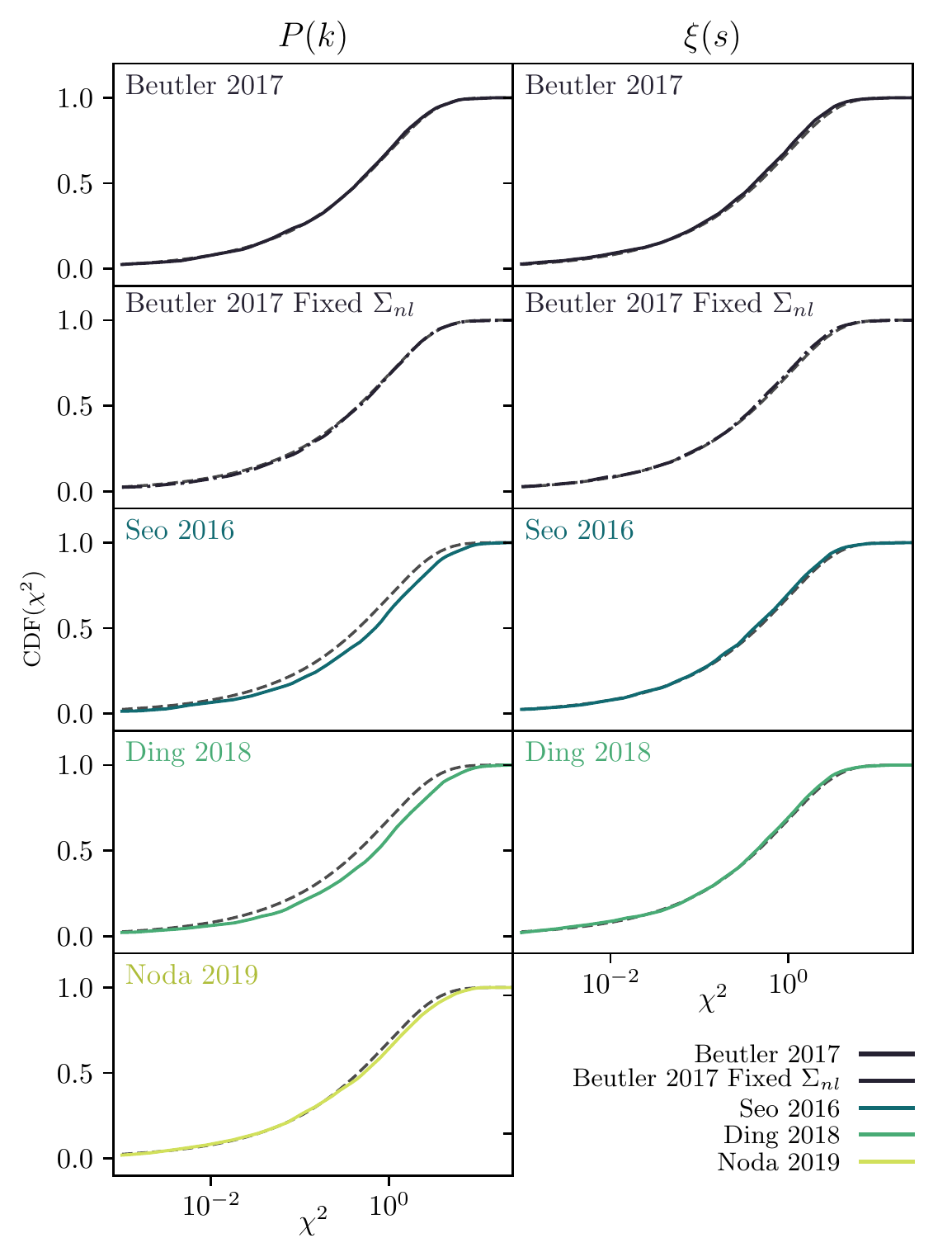}
    \end{center}
    \caption{The CDF distribution of the $\chi^2$ value for our thousand fits to the mocks, where the $\chi^2$ values are given as $(\Delta\alpha/\sigma_{\alpha})^2$. The dashed line shows the analytic $\chi^2$ CDF with one degree-of-freedom. Models with a CDF falling under the analytic distribution are under-predicting their uncertainty, and a model with errors perfectly representative of the scatter between mocks would lie exactly on the analytic $\chi^2(1)$ CDF.}
    \label{fig:chi2}
\end{figure}

As for our power spectrum tests we next look at the error on $\alpha$ recovered for each of our mocks and model combinations. As can be seen in \Cref{fig:xi_avg_summary2} the choice of model used changes the error slightly less compared to the result for the power spectrum fits, in accordance with the higher degree of correlation between results shown in \Cref{fig:xi_avg_individual_alphacomp}. Other trends are qualitatively the same as for the power spectrum; the {\beutler} model when $\Sigma_{nl}$ is allowed to vary generally gives larger errors, but does occasionally result in better errors than other models when noise enhances the BAO feature; it is hard to choose which of the {\ding} and {\seo} models reports the smallest error; and we see no correlation between the best-fit value of $\alpha$ and it's uncertainty for any models. Unlike for the power spectrum, we find no evidence that any of the models are under-reporting the uncertainties on the individual realisations --- the distribution of $\alpha$ constraints about the expected value is perfectly consistent with the $\chi^{2}$ distribution as seen in \Cref{fig:chi2}. 

\subsection{Comparison to Power Spectrum}

\begin{figure}
    \begin{center}
        \includegraphics[width=\columnwidth]{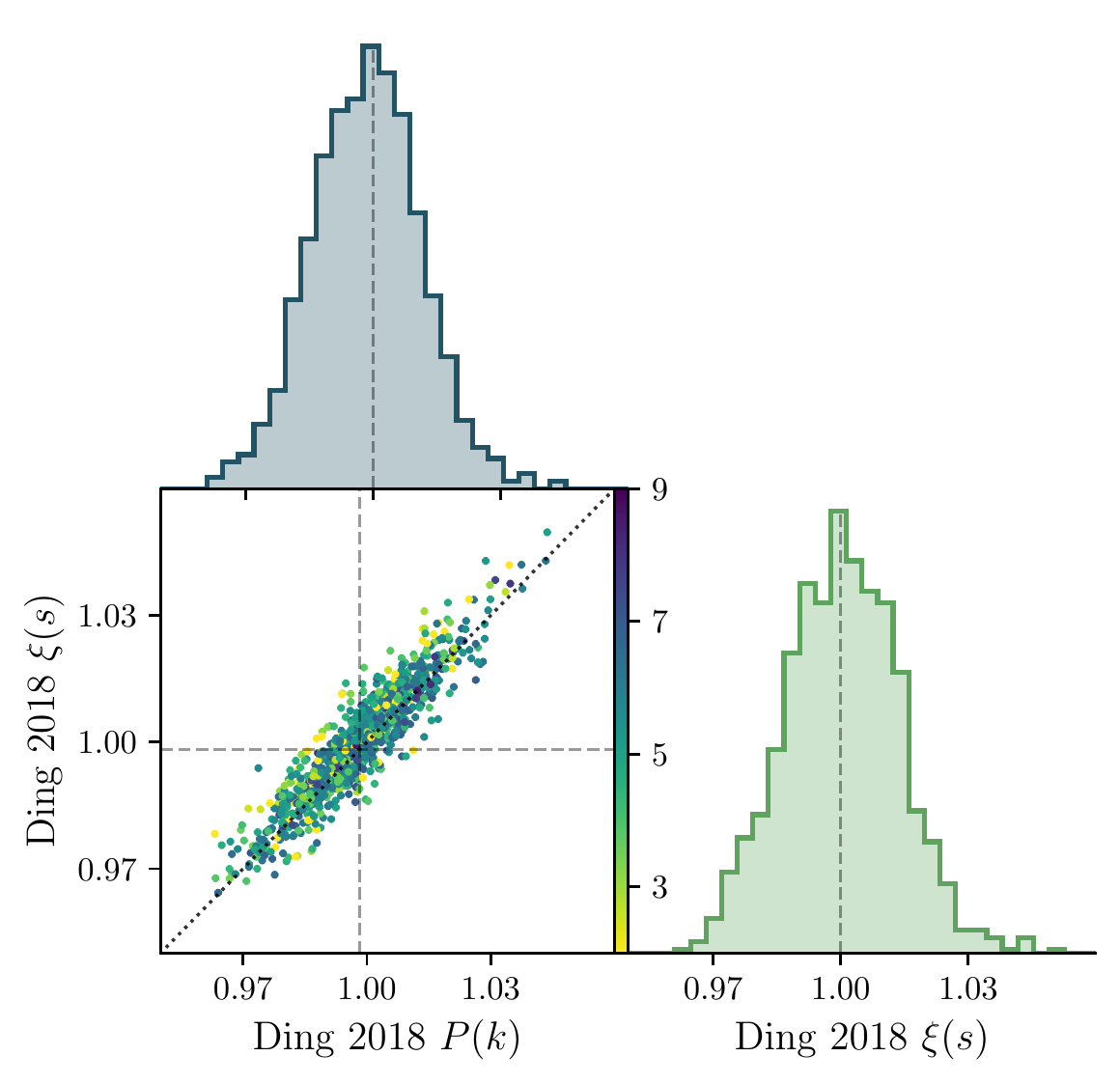}
    \end{center}
    \caption{$\alpha$ recovery from $P(k)$ and $\xi(s)$ using the {\ding} model for each of the 1000 mocks. The colourbar shows the detection strength in $\sigma \equiv \sqrt{\Delta\chi^2}$. All points give a correlation of $0.92$, rising to a value of $0.94$ when implementing a $5\sigma$ detection significance requirement.}
    \label{fig:pk_vs_xi_individual_alphacomp}
\end{figure}

In order to verify the correctness of our model transformations into real-space, we run a comparison using the {\ding} model on all thousand mocks, comparing the results from the power spectrum and correlation function fits. For ease of presentation, we do not show every model, allowing {\ding} to illustrate the correlation by itself. These results are shown in \Cref{fig:pk_vs_xi_individual_alphacomp}. Without any cuts on detection significance, we find a correlation of $0.92$. Implementing a $5\sigma$ cut, this rises to a correlation of $0.94$, agreeing with the correlations of roughly $0.9$ found in \citet{Sanchez2017} and $0.95$ from \citet{Anderson2014}. 

These results indicate a smooth implementation of the correlation function models, and would allow us to combine the two measurements into a single consensus result if we wished. As mentioned above, the correlation function models on average report less bias than their  power spectrum counterparts, at the cost of slightly increased uncertainty. These findings persist even when unifying the fitting ranges of the two methods.

\section{Consensus Measurements}
\label{sec:consensus}

In this work, we have provided a detailed study of 9 different models that can be used to measure the BAO feature in measurements of the galaxy power spectrum or correlation function. For all of these models (with a few tweaks) we are able to recover unbiased constraints on $\alpha$, but it is not clear which of the models is ``best''. For instance, one could be tempted to simply choose the model that reports the lowest error on average or try to combine all the models taking into account the correlations between the models measured from the simulation ensemble using for instance the Best Linear Unbiased Estimator (BLUES) method for correlated measurements (see e.g., \citealt{Winkler1981} and \citealt{Sanchez2017}).

However, there is an important caveat to this which is that such procedures are only suitable if the model with the smallest error is returning a representative value. In order to test this, as alluded to in Sections~\ref{sec:pk} and \ref{sec:xi}, we looked at the distribution of $\chi^{2}$ values from the post-reconstruction mocks for each of the models tested and compared these to the expected distribution for a single degree-of-freedom. The cumulative distributions for each model are shown in \Cref{fig:chi2}. As can be seen, the {\seo} and {\ding} models do not follow the expected distribution when applied to the power spectrum. These models underestimate the errors on some of the mocks. As such, simply taking the minimum error model could cause an underestimation of the error on the data, whilst naively combining the results even accounting for the correlations will cause an even larger underestimation. In the worst case scenario, such a combination would underestimate the error both due to the fact that some of the models individually under-report their uncertainties, and additionally because the measurements from the different models are in tension at the level of their quoted uncertainties.

\begin{figure}
    \begin{center}
        \includegraphics[width=\columnwidth]{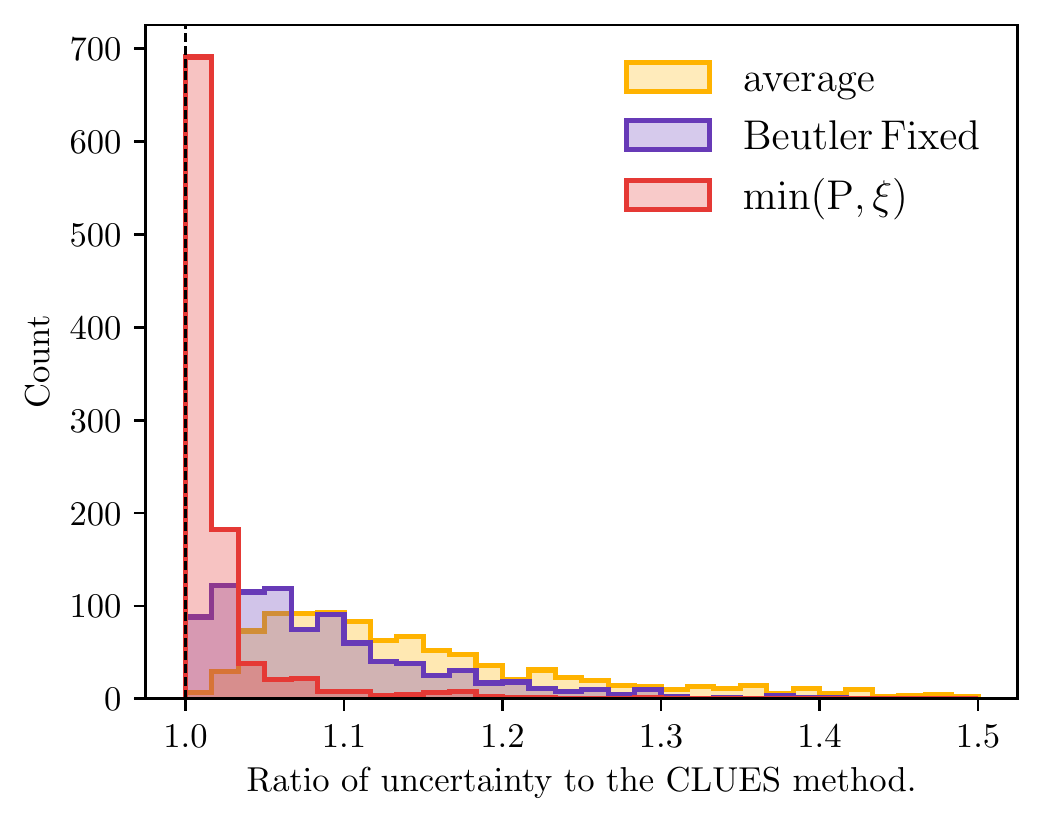}
    \end{center}
    \caption{Uncertainty for each mock when using different methods to come to a consensus result, compared to the uncertainty given by the `CLUES' procedure given in \Cref{sec:consensus} (inflating the errors and then combining $P(k)$ and $\xi$ measurements). `average' are obtained by integrating over the posteriors of all 9 models, `Beutler Fixed' is the result from fitting the power spectrum with fixed value of $\Sigma_{nl}$ (shown for comparison), and `min(P,$\xi$)' is when the smallest uncertainty is taken from either the power spectrum or correlation function fits \textit{after} error inflation. We see that, compared to the CLUES procedure, other methods provide larger uncertainty, although the `average' method could be viewed as the most conservative and does not require tuning the errors from individual models.}
    \label{fig:consensus}
\end{figure}

We tested a number of alternative different methods for combining the different models and results and ultimately settled on the following procedures. If, philosophically, we consider that differences between any of the models arise due to small modelling systematics, then averaging over the 9 posterior distributions will result in a conservative and unbiased estimate of the measurement constraint. Doing so, we obtained a typical error $\langle \sigma_{\alpha} \rangle=0.0134$, which is larger, for instance, than taking the power spectrum only result using the {\beutler} model with Fixed $\Sigma_{nl}$ ($\langle \sigma_{\alpha} \rangle=0.0122$, chosen because it closely matches the method used in the original BOSS-DR12 analysis). However, one could argue that this result is more robust because it accounts for model variations and variations between the correlation function and power spectrum results.

An alternative method we found which results in a smaller error than either this or the {\beutler} model, but still retains the expected distribution of constraints on a mock-by-mock basis is the following algorithm, which we label CLUES for convenience - an extension of BLUES with corrected uncertainty.

\begin{itemize}
    \item {First, we add an additional error component to the models, optimised to reduce the Anderson-Darling test statistic between the mock $\chi^{2}$ values and the chi-squared distribution. Only the {\seo}, {\ding} and {\noda} models returned a best-fit additional component that was non-zero corresponding to increasing the typical errors from these models by $22\%$, $25\%$, and $5\%$ respectively.}
    \item{After inflating the errors for these models, we then chose the model with the smallest error for each mock for the power spectrum and correlation function separately.}
    \item{ Taking that smallest error measurement for each mock, we compute the cross-correlation between the power spectrum and correlation function results.}
    \item{The two minimum-error measurements (for $P(k)$ and $\xi(s)$ after error inflation) are then combined using the empirical cross-correlation coefficient.}
\end{itemize} 
Philosophically, this procedure is akin to assuming that differences between different models fit to the same clustering statistic arise due to modelling systematics or under-estimation of uncertainties, but that the constraints from the power spectrum or correlation function differ due to their information content or data systematics such that there is benefit in combining them.

Overall, we find that the above procedure CLUES returns smaller uncertainties ($\langle \sigma_{\alpha} \rangle=0.0114$) than averaging over all models, using the {\beutler} model, or taking the measurement with the smallest error from either the power spectrum or correlation function (after error inflation where necessary). This is demonstrated in \Cref{fig:consensus}, where we plot the ratio of the errors against all these three alternatives. Crucially however, our combination remains completely unbiased both in terms of the mean and scatter of best-fit $\alpha$ values between mocks, and the distribution of mock $\chi^{2}$ values. There is only a small improvement performing a weighted combination of the power spectrum and correlation function constraints compared to simply taking the tightest of the two because the power spectrum results are better for $95\%$ of cases, however there is some still some small decrease in uncertainty, which is available at little cost given that most studies fit both of these statistics anyway.

The algorithm presented in this section applies well to a single parameter. Scaling the algorithm up to provide consensus measurements on multiple parameters (for example, $\alpha_\perp$ and $\alpha_\parallel$) would require generalising all 1D tests and comparisons into higher dimensions. A potentially viable methodology would be to compare each model against the $\chi^{2}$ distribution accounting for the covariance matrix of the parameters and using the correct number of degrees-of-freedom; and to replace selecting models based on the smallest error on $\alpha$ with the result that gives the smallest determinant of the parameter covariance matrix, as this would correspond to the smallest hyper-volume error ellipse in arbitrary dimensions.

As a final note, we caution that the exact results/numbers above will certainly vary for different datasets. Nonetheless, the detailed comparison of models, and our procedure to combine them into a robust consensus, could be readily tested on any current or upcoming dataset, especially considering the ease with which \textsc{barry} can produce all the necessary fits.


\section{Conclusions}
\label{sec:conclusion}

In this paper we present a comparison of state-of-the art models for fitting isotropic Baryon Acoustic Oscillations in the clustering of galaxies. This is made possible through a new python framework \textsc{Barry}, which is modular, open-source, and enables a robust comparison of the different models independent of details that may affect evaluation of their relative performance (such as calculation of the underlying smooth matter power spectrum, and sampling of the parameter posteriors). 

We have tested 9 different models (5 for the power spectrum and 4 for the correlation function) on the MultiDarkPATCHY mocks originally designed for analysis of the SDSS-III BOSS DR12 data. We are able to recover (after some modifications to the non-linear components of the {\noda} model) unbiased results with all models when fitting for the isotropic BAO feature, pre- and post-reconstruction and in either the power spectrum or correlation function formulations. In addition we highlight:
\begin{itemize}
    \item{The impact of fixing $\Sigma_{nl}$ for the model inspired by \citet{Beutler2017}. We found this significantly reduced fitting outliers when compared to other models. Allowing this to vary does result in improved constraints for some mocks, but this occurs when noise sharpens and enhances the BAO feature (as can be seen in the correlation between smaller $\alpha$ error and smaller values of $\Sigma_{nl}$). As such, the model reporting a smaller error may be undesirable.} 
    \item{Models that physically model the damping of the BAO feature (in particular the {\seo} and {\ding} models) produce tighter constraints, but risk underestimating the errors on the data and so must be applied with care.}
    \item{The claims made in \citet{Noda2019} with regards to improved BAO constraints using their BAO extractor method have been found to be caused primarily due to their fixing of free parameters. The extractor method applied with index mixing as per \citet{Noda2019} failed to achieve any fitting benefit when used alongside the {\ding} model, as the presence of the original power spectrum requires the same flexibility in the models as simply not using the extractor at all.}
\end{itemize}  

Overall, we also find higher model correlation in correlation function fits than in power spectrum fits, indicating that the choice of model is less important if working with the correlation function. Using standard fitting ranges for both $P(k)$ and $\xi(s)$ result in marginally lower uncertainty for the power spectrum models compared to correlation function models, even when accounting for potential under-reporting of uncertainty. 

In future work, we will extend \textsc{Barry} to include anisotropic BAO models and continue improving numerical efficiency to allow for further high-precision systematic tests of the next generation of large-scale structure surveys such as DESI \citep{DESI2016}, 4MOST \citep{4MOST2019} and Taipan \citep{DaCunha2017}. \textsc{Barry}, and all the results and code used in this work, are publicly available,\footnote{\url{https://github.com/Samreay/Barry}} and we plan to incorporate \textsc{Barry} into the analysis of the above surveys.

\section*{Acknowledgements}

We thank Florian Beutler for providing the SDSS DR12 mock power spectrum measurements and Cheng Zhao, Albert Chuang and Ashley Ross for the correlation function measurements. We also thank David Parkinson and Pat Scott for useful discussions. Computational tasks were performed on the SMP getafix cluster computer at the University of Queensland. 

The authors acknowledge support by  the  Australian  Government  through  the  Australian Research  Council's  Laureate  Fellowship  funding  scheme (project FL180100168). This research has made use of NASA's Astrophysics Data System Bibliographic Services and the \texttt{astro-ph} pre-print archive at \url{https://arxiv.org/}. Plots in this paper were made using the {\sc matplotlib} plotting library \citep{Hunter2007} and the {\sc ChainConsumer} package \citep{Hinton2016}. This research made use of Astropy,\footnote{\url{http://www.astropy.org}} a community-developed core Python package for Astronomy \citep{astropy:2013, astropy:2018}.


\bibliographystyle{mnras}
\bibliography{bibliography}







\bsp	
\label{lastpage}

\end{document}